\documentclass[final,3p]{elsarticle}

 \usepackage{graphics}
 \usepackage{graphicx}
\usepackage{caption}
\usepackage{float}
\usepackage{subcaption}
\usepackage{amsmath}
\usepackage{amssymb}
\usepackage{color}
\usepackage{soul}           
\usepackage{ulem}
\usepackage{tikz}
\usepackage{makecell}
\usepackage{algorithm}
\usepackage{algpseudocode}
\usetikzlibrary{shapes.geometric, arrows}

\tikzstyle{normal} = [rectangle, rounded corners, minimum width=3cm, minimum height=1cm,text centered, draw=black, fill=red!30,text width=10em]

\tikzstyle{robust} = [trapezium, trapezium left angle=70, trapezium right angle=110, minimum width=3cm, minimum height=1cm, text centered, draw=black, fill=blue!30,text width=4em]

\usepackage[T1]{fontenc}
\usepackage{hyperref}
\hypersetup{
colorlinks=true
}
\usepackage{epstopdf}
\usepackage{xcolor}

\biboptions{sort&compress}

\journal{arXiv}

\begin{document}

\begin{frontmatter}

\title{\textcolor{black}{Identification of fault frequency variation in the envelope spectrum \\in the vibration-based local damage detection in changing load/speed conditions}}

\author[label1]{Daniel Kuzio}
\author[label1]{Rados{\l}aw Zimroz}
\author[label2]{Agnieszka Wy\l oma\'nska}

\address[label1]{Faculty of Geoengineering, Mining and Geology, Wroclaw University of Science and Technology, Na Grobli 15, 50-421 Wroclaw, Poland }
\address[label2]{Faculty of Pure and Applied Mathematics, Hugo Steinhaus Center, Wroclaw University of Science and Technology, Wyspianskiego 27, 50-370 Wroclaw, Poland}


\begin{abstract}
\textcolor{black}{Local damage detection in bearings/gears is an important problem in machine condition monitoring. The diagnosis of such damage is based on the detection of impulsive and periodic signal. Both conditions should be checked as the identified fault frequency must be linked to the true value calculated for a given machine and operational conditions. In case of a complex machine (with many components), low rotational speed, poor signal-to-noise ratio (SNR), and some fluctuations in rotational speed, the precise estimation of the fault frequency is not a trivial task. One needs to find a segment of the signal with the appropriate length to find the fault frequency value. Selection of a long signal can improve the precision of frequency estimation; however, it can be difficult to find a homogeneous segment in the case of time-varying operating conditions (load/speed variation). In the case of early stage damage diagnosis, a poor SNR may influence the estimation error. Thus, finally, each measurement may provide different results, so a statistical analysis may be necessary.  A fault frequency distribution for a set of signals should be considered. Due to the factors mentioned above, the distribution of estimated fault frequencies values may be destroyed and thus misleading. In this paper, we present simulation studies to illustrate the problem and several industrial examples to prove the practical importance of our research topic. Finally, we propose a simple procedure to evaluate whether the speed (fault frequency) is constant or not, and if not, what is a possible source of these variations (true input speed changes or methods used for fault frequency estimation). The proposed algorithm uses the fault frequency estimation technique based on peak detection in the envelope spectrum and the methodology for statistical testing if the estimated fault frequency can be considered as a constant value. 
The procedure helps to decide whether frequency variation exists or not and what is a possible source of such variation that may suggest next steps in decision making. 
}
\end{abstract}



\end{frontmatter}

\section{Introduction}
Local damage detection in bearings/gears is an important problem in machine condition monitoring. Specifically, damage detection at an early stage \cite{Zhang2023,capdessus2000cyclostationary,combet2009optimal}, with possible variation in operational condition \cite{Liu20202323,Schmidt2021_169,schmidt2020methodology,Sun2021,Schmidt2021_158,Mauricio2022,Mauricio2020_144,Xu2022}, and in the presence of non-Gaussian noise \cite{borghesani2017cs2,Wodecki2021,cvb,hebda2020selection} is challenging. Most of the methods are based on the detection of the presence of an impulsive signal (in case of signal disturbed by Gaussian noise) or periodic nature of impulses. The second approach uses the so-called envelope analysis. It often requires prefiltering to improve the signal-to-noise ratio (SNR), amplitude demodulation, and spectral analysis of the obtained envelope \cite{randall2011rolling,Rai2016289,Zhao2022}.
It is obvious that in such a case the signal should be stationary and have approximately constant spectral content (constant fault frequencies). Under the assumption of possible speed fluctuation,  this approach requires a short segment of signal with homogeneous properties. However, if speed fluctuation exists, estimated fault frequency values for the next segments may be different. As the segment is short, resolution in the frequency domain may affect detection precision. Moreover, it may be the case that the spectral component is hardly detectable due to noise. Thus, as signal is periodic and contains some harmonics, one may detect first, second, third, etc. harmonic and after normalization (by harmonic number), the fault frequency may be calculated as an average value \cite{Urbanek_Jacek_Measurement_2011}. Even if the rotational speed is constant, due to segmentation, frequency resolution, and the presence of noise, many measurements may result with some distribution of fault frequency instead of one value. Unfortunately, small variations in rotational speed will provide a similar effect. 
If variation of speed is rather slow and follows some pattern, i.e. sine wave, the distribution of estimated frequency values may look like a uniform distribution.
All these factors may create a situation that is difficult to conclude if the fault frequency is exactly constant value or its fluctuation is related to speed variation or parameters of segmentation, SNR, etc.
Therefore, to investigate the problem, we propose a deep simulation study inspired by real cases. 
We consider 3 models of frequency behavior: constant value, some minor fluctuation of speed around mean value (equivalent to Gaussian distribution), and uniform distribution of speed from given range.
Then, we analyze several signals with different SNR levels. Additionally, an analysis was performed for several segment sizes. The size of the segment depends on the nature of the input speed.
Although local damage detection is a well-known problem and the speed of varying time has been recognized as a challenging problem, the perspective proposed in this paper has not been discussed in the literature. 

Obviously, as the topic is important and challenging, there are some interesting works related to bearing diagnosis under varying load. 
Liu et al. \cite{Liu20202323} proposed an iterative generalized demodulation with tunable energy factor (IGDTEF). Hou et al. \cite{Hou2021}  proposed a sparse time-frequency method for fault diagnosis without the needed speed information. 
Wang and Chu proposed a time-frequency representation-based order tracking algorithm. They were able to determine the informative frequency band and then resample the informative signal fault characteristic order spectrum \cite{Wang2019391}. Wang et. al \cite{Wang2019} employed polynomial chirplet transform to estimate the instantaneous rotation frequency (IRF) of bearings from vibration signals. On this basis, a maximum correlated kurtosis deconvolution-based envelope order spectrum is applied to detect bearing fault characteristic order. Ming et al. \cite{Ming2016367} developed a new fault feature extraction and enhancement procedure that consists of the combination of iterative envelope analysis and a low-pass filtering operation. They obtained the temporal envelope signal, which was next transformed to the angular domain by the computed order tracking, and the fault "frequency" was detected in the classical way in the envelope spectrum. 
In \cite{Borghesani201323} a new procedure called RS-SES (reverse sequence squared envelope spectrum procedure) was proposed that combines squared envelope spectrum and computed order tracking was proposed in \cite{Borghesani201323}.

Zhao et al. \cite{Zhao2016109} proposed a novel method for compound faults detection of rolling element bearing based on the generalized demodulation (GD) algorithm. Zhao et al. \cite{Zhao2015} proposed a method for identifying faulty bearing characteristics based on a combination of instantaneous dominant meshing multiply (IDMM) and empirical mode decomposition (EMD). 
All of these techniques have been tested successfully; however, in this paper, as mentioned, the perspective is different. We do not use order spectrum or resampling (tacho or vibration-based).

The motivation for this research comes from heavy-duty mining machines. They work with varying load/speed conditions \cite{Chaari_2012_SV} and have a significant speed reduction (even from 1000 rpm to 5 rpm) because they use multi-stage gearboxes. This means that selection of appropriate signal length for fault frequency estimation is really not a trivial task - it should be long as one plays with low frequency components, but it should be short as speed is varying. Low speed also means significantly smaller dynamics of kinematic pairs, so detection of spectral components is also much harder. Such a problem may also appear, for example, in wind turbines and has probably a wider impact there. 
In the proposed simulations, we present the influence of given factors on the final estimation of the frequency of the fault. Finally, we present several industrial examples with mentioned speed variation. 

The rest of the paper is organized as follows. In the next section, we present the problem of variation of operating conditions and discuss the influence of a speed change on the detection of a fault frequency. Some examples of real data are provided to prove that the problem exists for many types of machine. Then, the link between the input speed distribution and the detected fault frequency distribution is presented as a block diagram with highlighted key steps that influence the final results (segmentation, SNR, spectral component detection). In Section \ref{sec3} basic theory of the signal model used for simulations, the methods of signal processing and statistical analysis of the final results are recalled. Finally, a procedure is proposed to estimate the distribution of fault frequency. The experimental part is covered by Sections \ref{simul} and \ref{real}, respectively, where simulated and real data are analyzed. The last section concludes the paper.


\section{\textcolor{black}{Problem formulation}}\label{pr_formulation}
\begin{figure}[H]
\centering
\hspace*{-2.2cm}
    \includegraphics[width = 8cm]{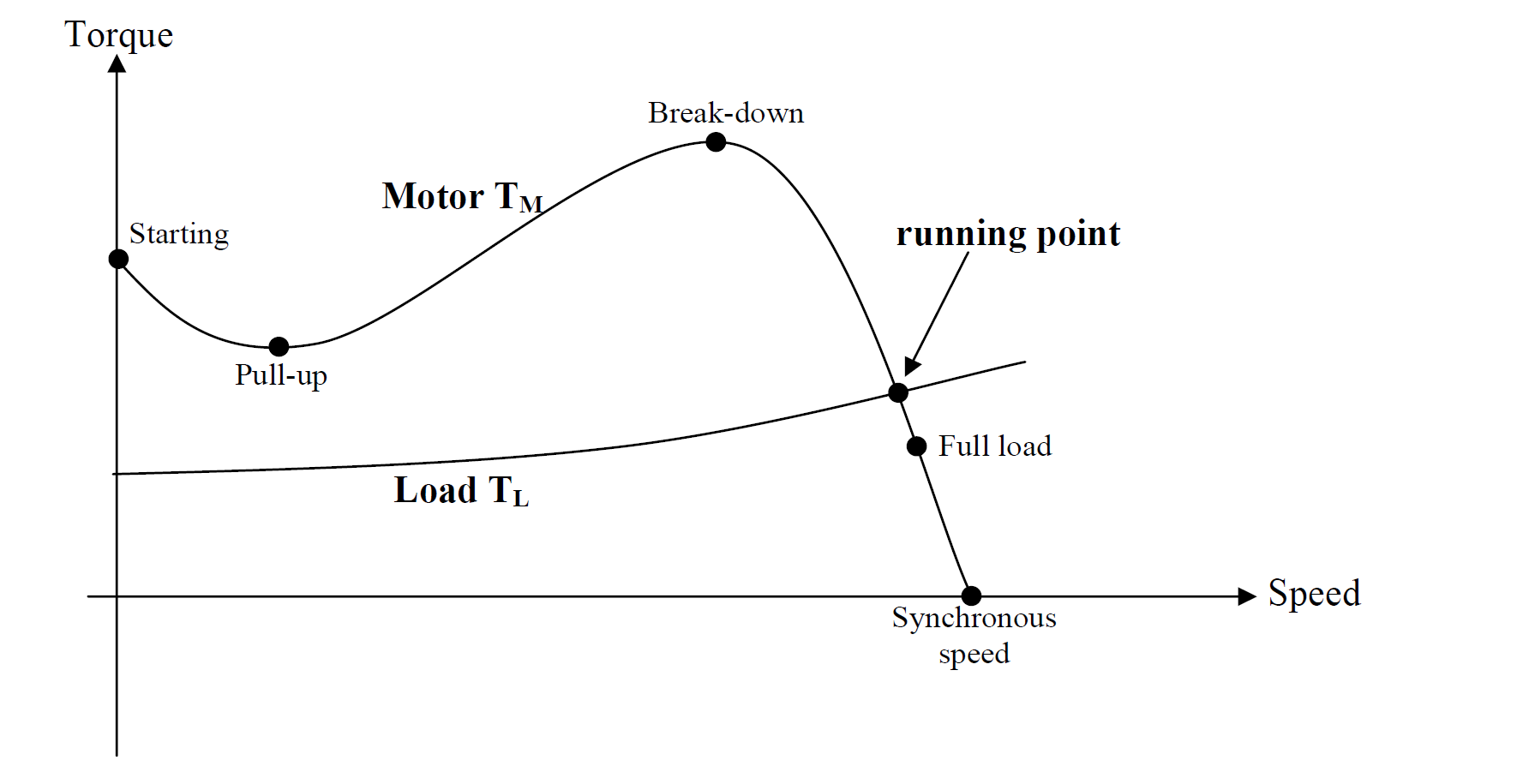}
    \caption{Torque-speed relationship in electric motor}
    \label{Fig1_motor}
\end{figure}

Non-stationary operating conditions are directly related to load or speed changes. In some cases, the change in external load directly influences the speed, as shown in Fig. \ref{Fig1_motor} (see the change in the running point). The reason for load/speed variations may be different, and, moreover, the character of variation and its range may depend on the machine design, technological process, etc. From signal processing point of view, variation of the speed may be deterministic or random, slow or fast, may follow specific pattern or function (liner, exponential, sine wave, etc.), or combination of mentioned above. 

\begin{figure}[H]
\centering
\hspace*{-1 cm}
    \includegraphics[width = 16cm]{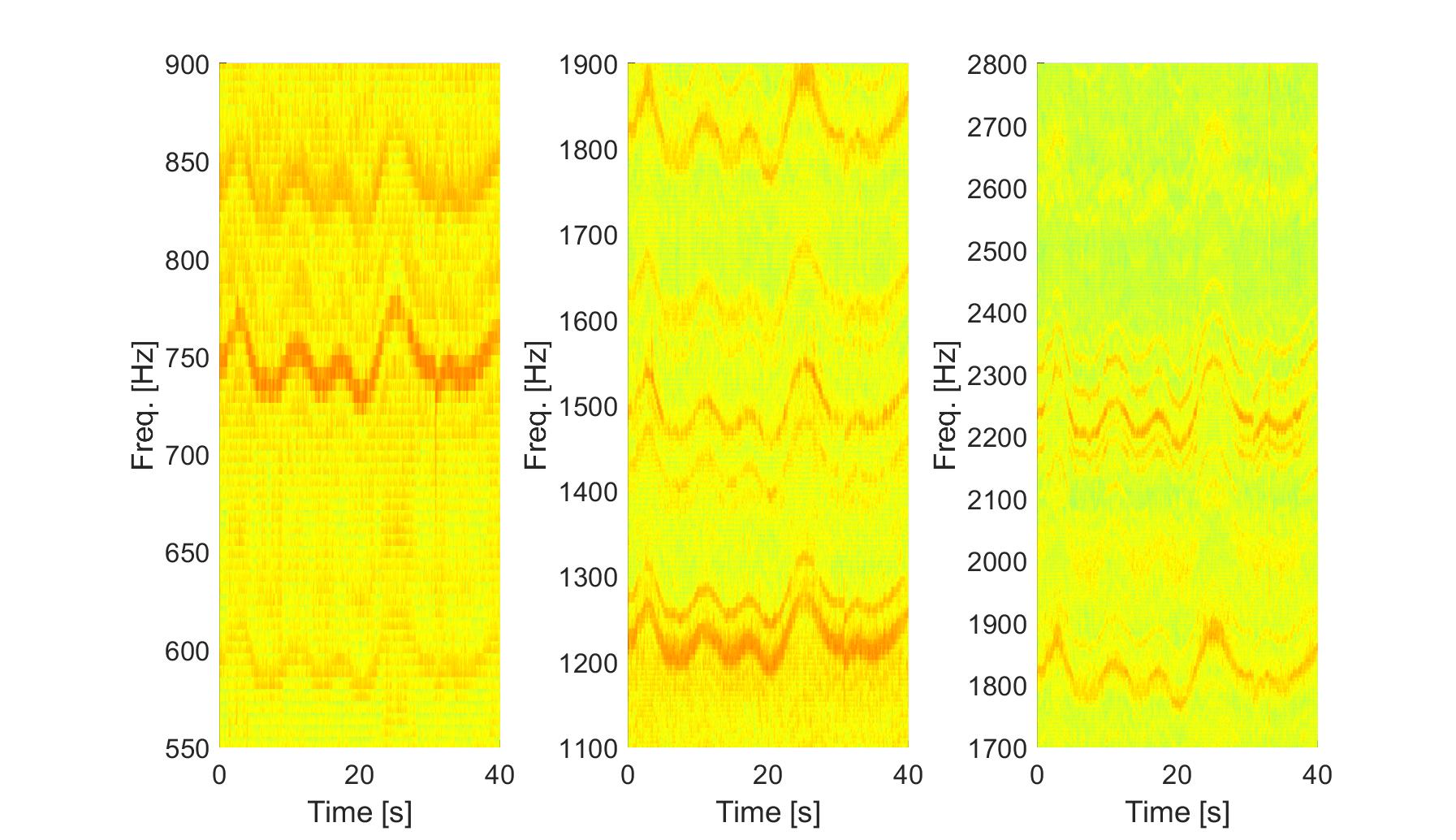}
    \caption{Time-frequency representation of wind turbine gearbox: a) around first mesh harmonics (750Hz), b),c) second and third harmonics }
    \label{Fig2_spectrogram}
\end{figure}

\begin{figure}[H]
\centering
    \includegraphics[width = 10cm]{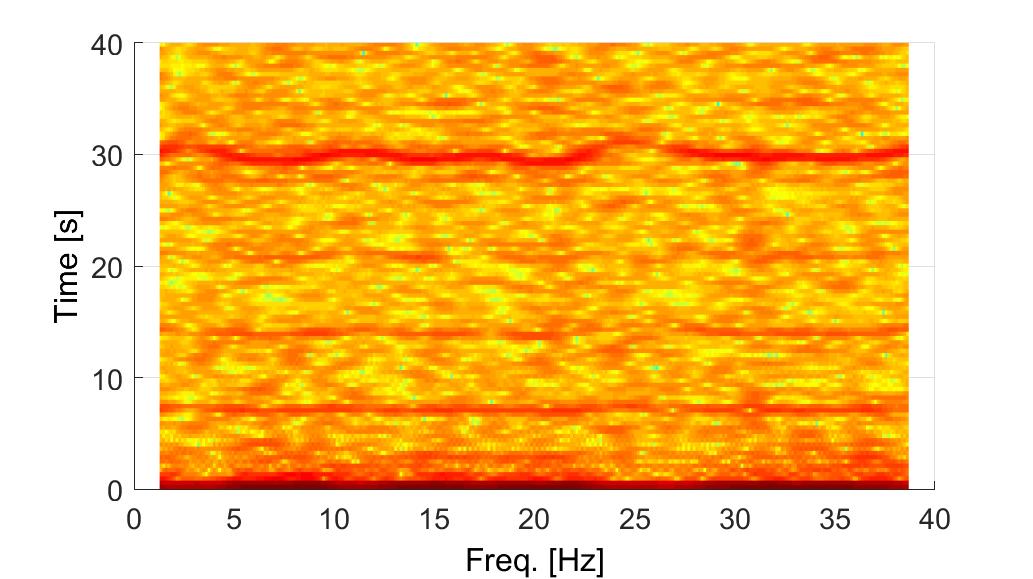}
    \caption{Spectrogram of envelope signal from wind turbine } 
    \label{Fig3_envelope}
\end{figure}

\textcolor{black}{Change of load and/or speed condition can affect the spectral content of the vibration signal. Variation of load may influence the amplitude of the spectral components, variation of speed may influence the value of the frequency, as well as the amplitude of the components, as it may interact with natural frequencies (resonances and anti-resonances). In Fig. \ref{Fig2_spectrogram} one can see that the variation of the first, second, and third mesh frequencies in the exemplary vibration of a gearbox is significant. Moreover, after demodulation around mesh frequency, the variation of side-band frequencies is still significant; see Fig. \ref{Fig3_envelope}.}

\begin{figure}[H]
\centering
    \includegraphics[width = 8cm]{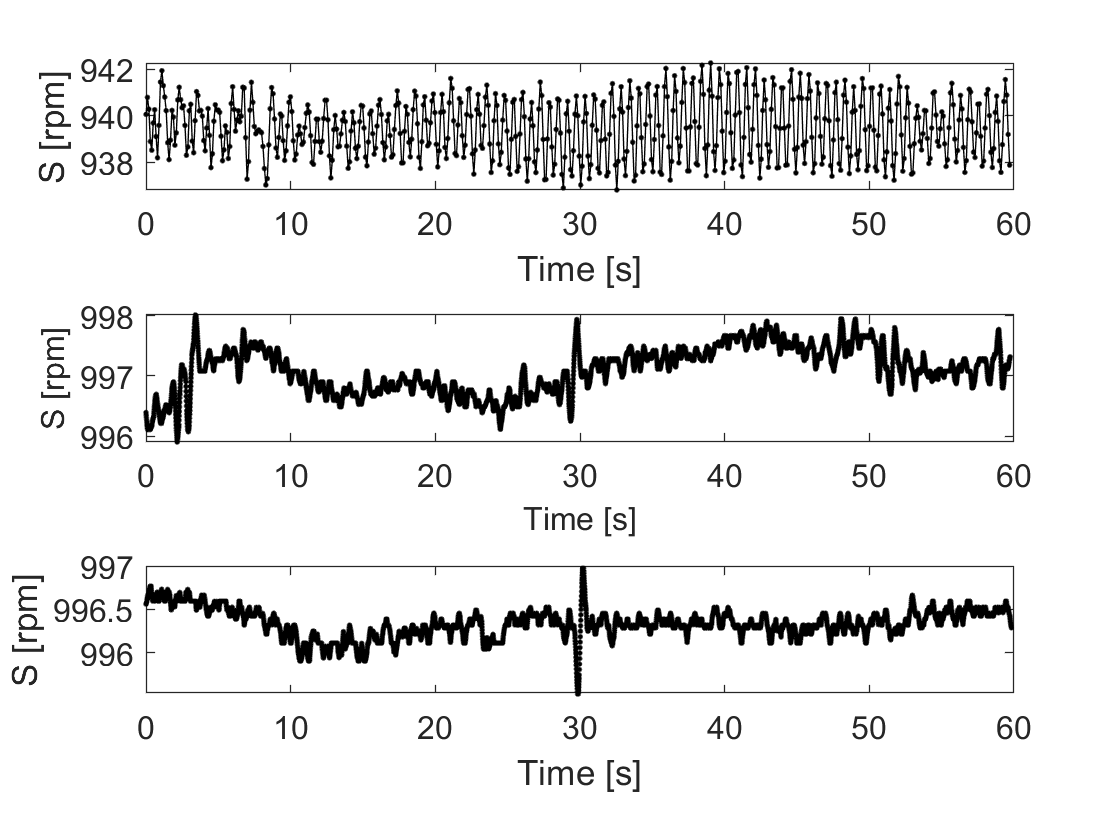}
    \caption{Examples of speed profile for belt conveyors }
    \label{speedprofileA1}
\end{figure}

\begin{figure}[H]
\centering
    \includegraphics[width =14cm]{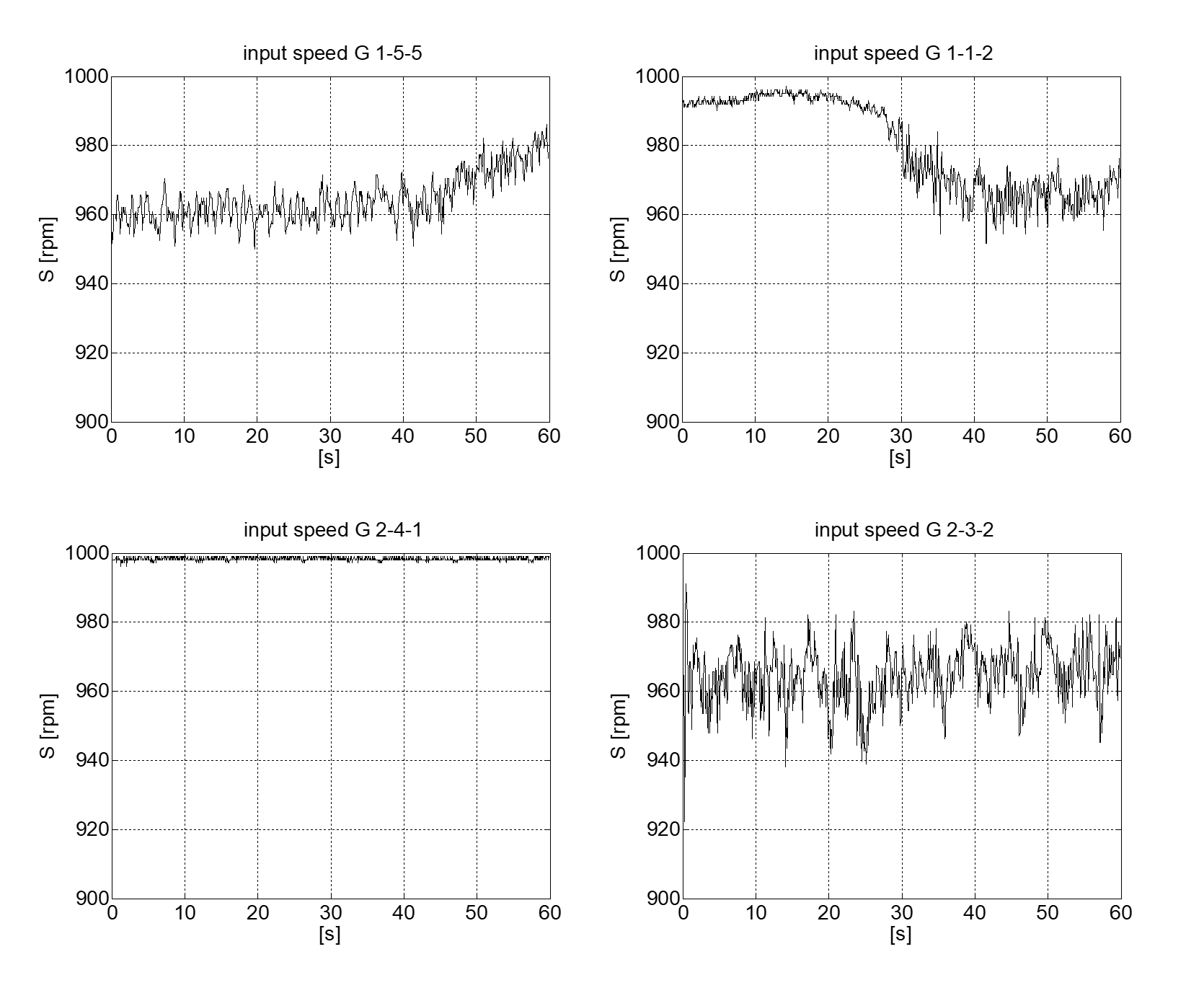}
    \caption{Examples of speed profiles for bucket wheel excavator}
    \label{speedprofileA}
\end{figure}

\textcolor{black}{Speed variation is phenomenon that frequently appears in real life. Some exemplary load/speed profiles from a belt conveyor \cite{Kruczek_poisson} and a bucket wheel excavator \cite{COMBET20091382}  used in the raw material industry can be found in Figs. \ref{speedprofileA1} and \ref{speedprofileA}, respectively. For more examples related to wind turbines, we refer the readers to \cite{Urbanek_Jacek_Measurement_2011}, for the packaging machine, we recommend \cite{COCCONCELLI2012667}.}

{The examples presented above are evidence of speed variation in real machines. Assuming the random nature of speed variation, we propose to analyze the probability density function (PDF) to describe the character of speed variation. It should be mentioned that we consider three cases: no variation, i.e. constant speed (values are presented in Fig. \ref{Diagram1} panel (a) and the corresponding PDF in panel (b)) and uniformly or Gaussian distributed variation. In general, we are talking about speed fluctuation, not changes in a very wide range of speed. However, the measured signal divided into dozen segments without overlap (see Fig. \ref{Diagram1} panel (c))  will provide a vector of fault frequencies with some variation (see Fig. \ref{Diagram1} panel (f)). Each segment is analyzed separately using envelope spectrum (see Fig. \ref{Diagram1} panel (d)) and provides a single value of the estimated fault frequency (see Fig. \ref{Diagram1} panel (e)). Assuming some randomness in the estimated values, one may use statistical analysis of detected fault frequencies for all segments (see Fig. \ref{Diagram1} panel (g)). Ideally, the PDF shape should be the same in the output as was introduced in the input of the procedure (see Fig. \ref{Diagram1} panel (a)). Unfortunately, due to the mentioned factors (variation in speed, segment length, and frequency resolution), a significant difference between the input and output distributions may be observed. A distortion of the distribution could lead to inappropriate decisions regarding the variation of the fault frequency.
It is important to mention that, assuming a normal or uniform distribution of the fault frequency, we can also get a different PDF  than the theoretical one (see Fig. \ref{simul_density}).



\begin{figure}[H]
\centering
    \includegraphics[width=14cm]{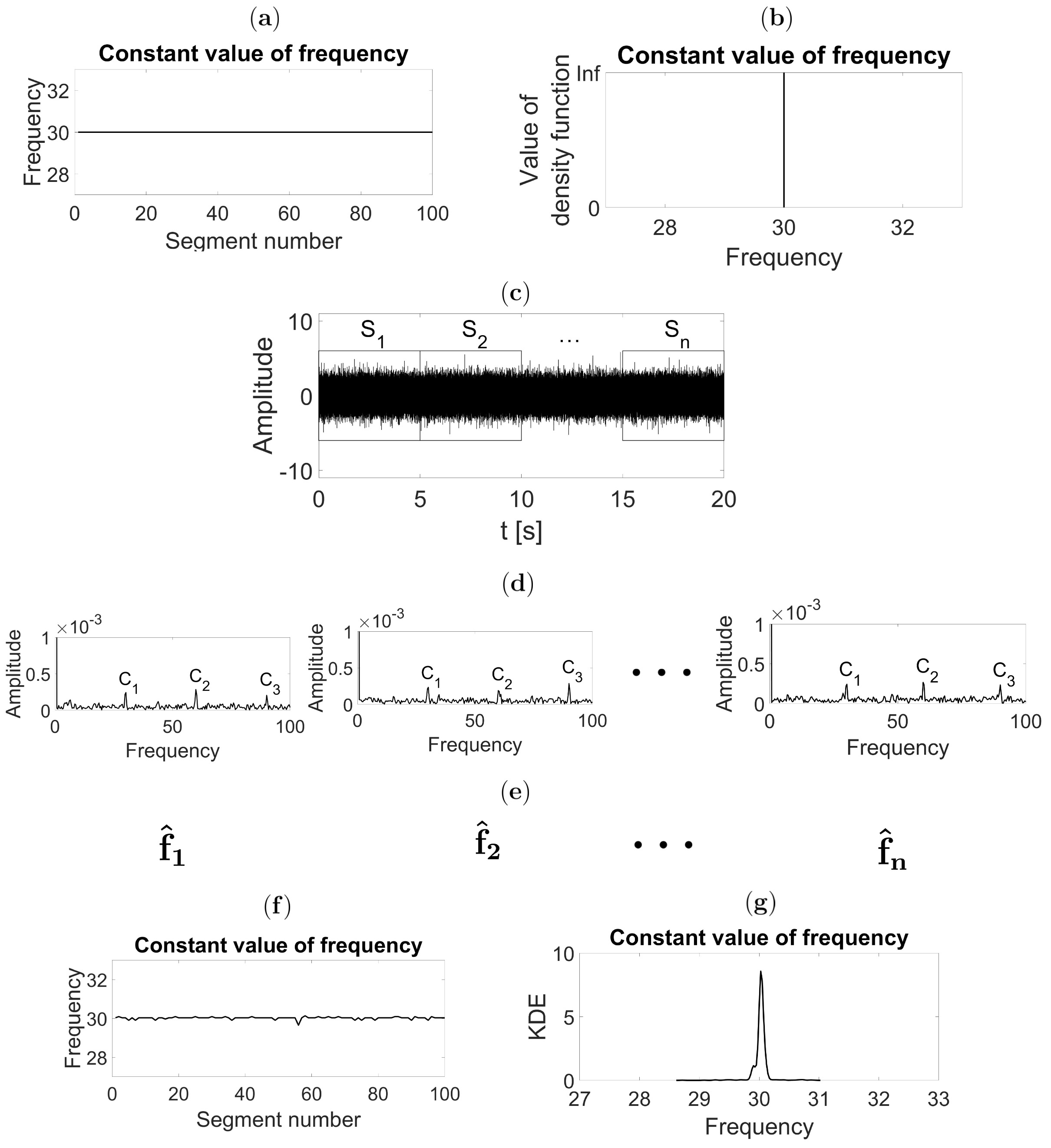}
    \caption{A reason for speed fluctuation in envelope spectrum}
    
    \label{Diagram1}
\end{figure}


\section{\textcolor{black}{Theoretical background}}\label{sec3}
\subsection{{Model of the signal}}\label{subsec3_1}
We assume that the signal model $x(t)$ consists of two main components: $x_1(t)$ and $x_2(t)$, which form an integrated mixture: $x(t)=x_1(t)+x_2(t)$, where
\begin{itemize}
    \item signal $x_1(t)$  is a Gaussian white noise with zero mean and unit variance,
    \item signal $x_2(t)=\text{ACI}\cdot gauspuls(t_{c},cf_{c},bw_{c})$, which is a signal of cyclic impulses associated with the local fault.  
\end{itemize}
In our case, we consider five different values of the amplitude of cyclic impulses (ACI), namely $\{1, 1.5, 2, 2.5, 3\}$. The cycle length, that is, the distance between impulses ($t_c$) is equal to 1/f [s], which corresponds to the fault frequency $f$ (in $Hz$). The fault frequency is related to inner race damage in the bearings, and in our case it is a random variable from a given distribution (or in the ideal case it is just a constant value $30 Hz$). The carrier frequency of the cyclic impulses $cf_c$ is set at $2.5 kHz$ (a bearing structural resonance). The fractional bandwidth $bw_{c}$ is chosen from the uniform distribution in the interval $[0.4,0.5]$. We want to emphasize that an influence of jitter in bearing vibration was not considered here, but it will also contribute to frequency variation.

\subsection{\textcolor{black}{Classical envelope analysis - recall}}\label{estimation}
As mentioned, to detect damage in rotating machines, the envelope spectrum (ES) is used. The purpose of ES is to measure the periodicity of the envelope, a signal that is obtained after the amplitude demodulation procedure. Before demodulation, one should ensure appropriate frequency band selection, perform bandpass filtering for the selected band, and then use the Hilbert transform (HT) to determine the envelope signal (Matlab function hilbert.m is used). 
The Hilbert transform of the real signal $x(t)$ can be computed using the following formula: 
\begin{equation}
\label{eq:Hilbert}
h(t)=H\{x(t)\}=\frac{1}{\pi }\int \limits_{-\infty}^\infty {\frac{x(\tau )}{t-\tau }\hbox {d}\tau }. 
\end{equation}
The h(t) is complex and to detect periodicity of h(t), the magnitude of the analytical signal is used:
\begin{equation} 
\label{eq:Hilbert_signal}
x_{ht} (t)=\left| x(t)+j \cdot h(t)\right| =\sqrt{x^{2}(t)+h^{2}(t)}. 
\end{equation}

Finally, spectral analysis (power spectral density estimate via Welch's method - pwelch.m in Matlab) is applied:

\begin{equation}
\label{eq:Hilbert_spectrum}
ES(f)=\left| \hbox {PSD}(x_{ht} (t))\right| . 
\end{equation}

\textcolor{black}{The ES(f) is a vector of amplitudes of spectral components associated with the vector of frequencies. The theoretical value of the frequency is known, as it could be calculated on the basis of rotational speed and some parameters of machine design \cite{randall2011rolling}. However, as mentioned above, the instantaneous speed could have some variation. Thus, the location (in the frequency domain) of the spectral peak (max value) could fluctuate as well. To find an exact value of the fault frequency, for given value of speed, some frequency range around theoretical fault frequency is pre-selected, then the maximum value is searched. The maximum value of the amplitude is associated with a certain frequency, and this value is considered an estimated fault frequency. Depending on the frequency resolution and the value of the fault frequency, different values of the fault frequency detection error may be obtained. Thus, knowing that there are also harmonic components in the spectrum, we repeat this procedure for the second and third components, and after normalization by harmonic order, we may average these frequencies to decrease the estimation error. In this paper, we use three harmonics (fundamental, second, and third) to estimate the fault frequency for a given signal. More harmonics in theory may provide better results; however, amplitude of the higher component is usually smaller, so the SNR is worse and, in fact, it may be an additional source of error. }
\subsection{Statistical measures used for validation of the proposed methodology}
  The mean squared error (MSE) is used here as a model evaluation metric. The MSE of a model with respect to a test set is the mean of the squared prediction errors over all instances in the test set. The prediction error is the difference between the true value and the predicted one. In our case, we use MSE to measure the difference between the estimated $\hat{f}$ and the true value of $f$ of the fault frequency \cite{ref1}:
\begin{equation}
MSE_n = \frac{1}{n}{\sum\limits_{i=1}^n (f_i-\hat{f})^2},
\end{equation}
where $f_i$ is the true target value, $\hat{f}$ is the predicted value and $n$ is the number of estimated frequencies. This number is equal to the number of signals analyzed or the number of segments for which the fault frequencies are estimated.

For the evaluation of the proposed methodology, we also use the kernel density estimator (KDE), which can help measure the distance between the distributions of the true and estimated fault frequencies.
Let $\hat{f}_1,\hat{f}_2,\cdots, \hat{f}_n$ be independent observations of the estimator of the fault frequency. Formally, in this case, KDE can be expressed as \cite{kde}:
\begin{equation}
    \hat{p}_n(f) = \frac{1}{nh^d}\sum\limits_{i=1}^n K\left(\frac{f-\hat{f}_i}{h}\right),
\end{equation}
where $K(\cdot):\mathbb{R} \to \mathbb{R}$ is a smooth function called the kernel function, and $h>0$ is the smoothing bandwidth that controls the amount of smoothing (in this paper, we use the MATLAB ksdensity.m function with Scott's rule for kernel bandwidth: $h=1.06\cdot \hat{\sigma}\cdot n^{-\frac{1}{5}}$, where $\hat{\sigma}$  is the sample standard deviation of the considered sample). 
\subsection{Chi-squared test}\label{def_chi}
A chi-squared 
test is used here to verify whether the variance of a sample $\hat{f}_1,\hat{f}_2,\cdots, \hat{f}_n$ (being  realizations of the estimator of $f$) is equal to some specified value. In the considered case, it is a value equal to the sample variance of the estimated fault frequencies obtained for the simulated signals with constant $f$. In the description of the test presented below, the tested value is denoted as $\sigma^2$. The details of the chi-squared  test are as follows \cite{stat2}:
\begin{align*}
&H_0: &&\hat{\sigma}^2 = \sigma^2\\
&H_1: &&\hat{\sigma}^2 > \sigma^2\\
&\textit{Test statistic}: &&T = \dfrac{(n-1)\hat{\sigma}^2}{\sigma^2}\\
&\textit{Significance level:} &&\alpha\\
&\textit{Critical region (reject the null hypothesis):} &&T>\chi^2_{1-\alpha, n-1},\\
\end{align*}
where $\sigma^2$ is the tested variance, $\hat{\sigma}^2$ is a sample variance of $\hat{f}_1,\hat{f}_2,\cdots, \hat{f}_n$, and $\chi^2_{1-\alpha,n-1}$ is the critical value of the chi-square distribution with $n-1$ degrees of freedom for the significance level $\alpha$. 

\subsection{Considered distributions of fault frequency}
In this paper, we consider three possible distributions of fault frequency $f$. The first case corresponds to the constant value; i.e. this is the one-point distribution (here we assume $f=30 Hz$). Two other cases correspond to uniform and normal distributions. 

We assume that the random variable $F$ has a uniform distribution in the interval $[a,b]$, $F\sim U(a,b)$ if its PDF $p(\cdot)$ and the cumulative distribution function (CDF) $P(\cdot)$ are as follows, respectively \cite{forbes2010statistical}: 
\begin{align*}
&p(f) = \frac{1}{b-a}, \hspace{7mm}a\leq f\leq b,\\
&P(f) = 
\begin{cases}
0, &~~f<a,\\
\frac{f-a}{b-a}, &~~a\leq f\leq b,\\
1, &~~f>b.
\end{cases}
\end{align*}


We assume that the random variable $F$ has normal (Gaussian) distribution with parameter $\mu$ and $\sigma^2$, $F\sim N(\mu,\sigma^2)$ if its PDF $p(\cdot)$ and CDF $P(\cdot)$ are as follows, respectively, \cite{forbes2010statistical}:
\begin{align*}
&p(f) = \frac{1}{\sigma\sqrt{2\pi}}e^{-\frac{1}{2}\left(\frac{f-\mu}{\sigma}\right)^2}, &\hspace{-3.5cm}f \in \mathbb{R},\\
&P(f) = \frac{1}{2}\left[1+\textit{erf}\left(\frac{f-\mu}{\sigma\sqrt{2}}\right)\right], &\hspace{-3.5cm}f \in \mathbb{R}.
\end{align*}
where ${erf}(x)=\frac{2}{\sqrt{\pi}}\int_0^x e^{-t^2}dt$ 
is the error function, \cite{erf}, $\mu$ is the location parameter (mean), and $\sigma$ is a scale parameter (standard deviation).

It is worth mentioning that even if the range of considered distribution is unlimited (i.e., like in the Gaussian case), in the simulation study presented in the next sections, the parameters are selected in such a way that the PDF takes values close to zero for arguments higher than $\mu+3\sigma$ and lower than $\mu-3\sigma$.  
\subsection{Signal-to-noise-ratio (SNR)}\label{def_SNR}

The SNR is defined as follows \cite{SIJBERS19961157}:
\begin{eqnarray}\label{snrr}
    SNR = \frac{\overline{A_s^2}}{\overline{A_N^2}},
\end{eqnarray}
where 
\begin{itemize}
\item $\overline{A_s^2}$ is the sample mean of the squared amplitudes of peaks in the envelope spectrum related to the main fault frequency, its second and third harmonics (see Section \ref{estimation} for more details about the selection of number of used harmonics); 
\item $\overline{A_N^2}$ is the sample mean of squared amplitudes (up to the third harmonic) of the envelope spectrum without peaks mentioned above.
\end{itemize}

 
\subsection{{Procedure for fault frequency variation identification }}


In this section, we describe the proposed procedure used to identify the frequency variation of the fault. The corresponding block diagram is presented in Fig. \ref{block_diagram}. We proceed as follows:
\begin{itemize}
    \item First, we load the real signal and divide it into $n$ (non-overlapping) segments of a given length. In this paper, we consider segments of lengths corresponding to $\{0.5s, 1s, 2s, 5s,10s\}$. For each segment length, we separately repeat the steps described below. 
    \item For each segment, we calculate the ES and finally we estimate the fault frequency according to the procedure described in Section \ref{estimation}. In consequence, we obtain a set of fault frequencies $\hat{f}_1,\hat{f}_2,\ldots, \hat{f}_n$ and consider them as a random sample.
    \item For each segment we calculate the SNR, see Section \ref{def_SNR}. Then, we take the average of the SNR values. This step is crucial to calculate the threshold used in the procedure; see Fig. \ref{Threshold block diagram} for more details and the description given in the following part of this subsection.
    \item We simulate signals (according to the model given in Section \ref{subsec3_1}) of the same length as the analyzed segment for the fault frequency $f=30Hz$. For each considered ACI we simulate $1000$ signals. 
    \item  For each simulated signal corresponding to a given ACI we calculate the SNR (see Section \ref{def_SNR})  and calculate the average of the values obtained. In consequence for a given ACI we obtain one averaged SNR.
    \item We identify the ACI for which the averaged SNR is closest (in absolute value) to the SNR of the real signal. 
    \item To calculate the threshold, we take the simulated signals with ACI identified in the previous step. The procedure to calculate the threshold is presented in Fig. \ref{Threshold block diagram}. Moreover, for the simulated signals with ACI identified in the previous step, we estimate the fault frequencies and their sample variance that we denote as $\sigma^2$. 
    \item For the vector $\hat{f}_1,\hat{f}_2,\ldots, \hat{f}_n$ calculated for the real signal, we calculate the sample variance. We denote it by $\hat{\sigma}^2$. 
    \item In the next step, we re-scale the calculated sample variance by the factor $\left(\frac{\hat{f}_{real}}{\hat{f}_{simul}}\right)^2$, where $\hat{f}_{real} = \frac{1}{n}\sum\limits_{i=1}^n \hat{f}_{i}$ for the estimated frequencies of the real signal and similarly $\hat{f}_{simul} = \frac{1}{n}\sum\limits_{i=1}^n \hat{f}_{i}$ for estimated frequencies of the simulated signal with selected value of ACI. It should be highlighted that the re-scaling could be omitted but in that case there is need to simulate all required signals (with different ACIs, see the steps above) with the same fault frequency as in real signal. 
 \item We check if the re-scaled variance of estimated fault frequencies of the real signal is higher than the corresponding threshold. If not, we stop the procedure and identify the fault frequency of the signal as a constant value. Otherwise, we check if the re-scaled variance of $\hat{f}_1,\hat{f}_2,\ldots,\hat{f}_n$ is significantly higher than the tested value $\sigma^2$ by performing the chi-squared test (see Section \ref{def_chi}). More precisely, we calculate the test statistic (denoted as $T$, see details of the chi-squared test) and compare it with the critical value $\chi^2_{1-\alpha,n-1}$ at a given significance level $\alpha$. 
\item If hypothesis $H_0$ is rejected, that is, when the test rejects the hypothesis of equal variances,  we identify the distribution of frequency by comparing the KDE of the estimated fault frequencies for the real signal with the theoretical PDF (here uniform or Gaussian). 
    
\end{itemize}

\begin{figure}[H]
\centering
    \includegraphics[width = 14cm]{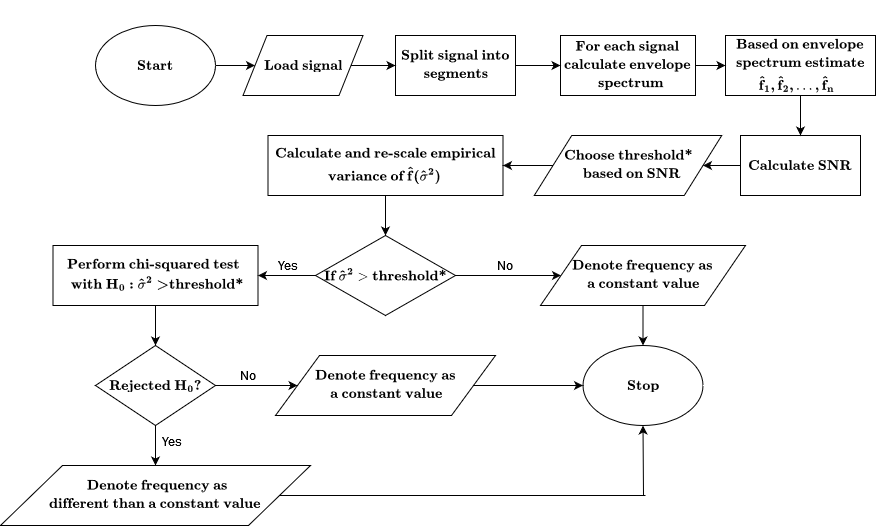}
    \caption*{*threshold -- block diagram for calculation is shown in Fig. \ref{Threshold block diagram}.}
    \caption{Block diagram of the procedure for identification of fault frequency distribution }
    \label{block_diagram}
\end{figure}

In the above procedure presented, the threshold value is needed. The threshold identification diagram is presented in Fig. \ref{Threshold block diagram}. In our procedure, the threshold is equivalent to the sample variance of the estimated fault frequencies for simulated data with selected ACI. 
 
At first, we load the $1000$ simulated signals with the identified ACI. The next step is to calculate the ES for each simulated signal. Then, on the basis of ES, we estimate $\hat{f}$ for all signals. As a consequence, we obtain $1000$ estimated values of the fault frequency.
Next, we calculate the sample variance of the estimated frequencies. Finally, we denote the already calculated variance as the threshold.

\begin{figure}[H]
\centering
    \includegraphics[width = 12cm]{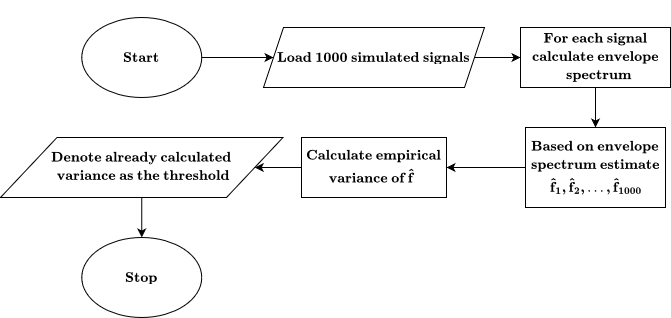}
    \caption{Block diagram for calculating threshold 
    }
    \label{Threshold block diagram}
\end{figure}

\section{Simulated signals analysis}\label{simul}

In this section, we analyze the simulated data to demonstrate the efficiency of the proposed methodology. Moreover, at the end of this section we present the table with the derived thresholds (for different segment lengths and ACI) that can be useful for the real signals analysis; see Table \ref{Thresholds_table}.

\begin{figure}[H]
\centering
    \includegraphics[width = 15cm]{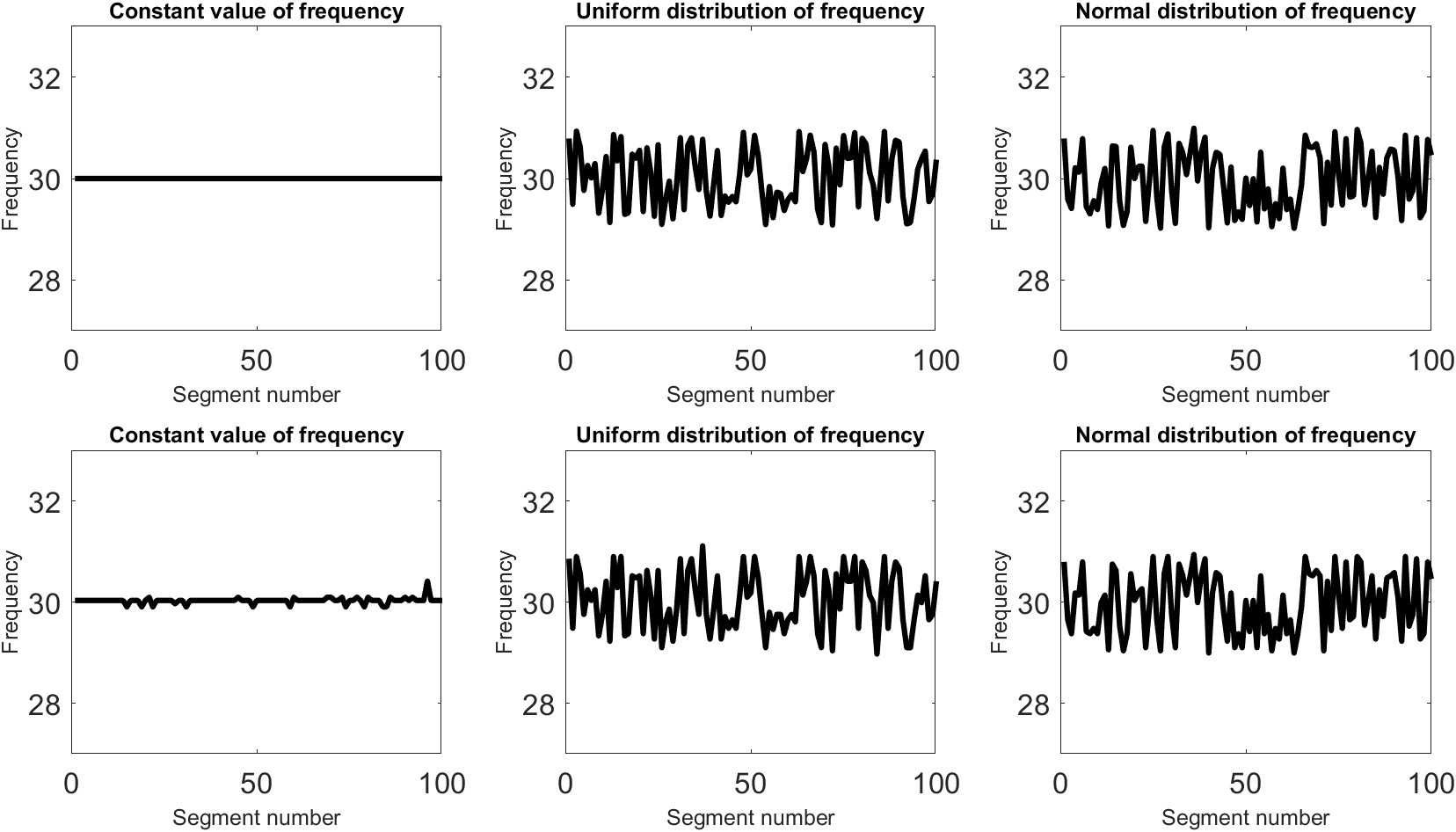}
\caption{Top panels: 100 simulated fault frequencies for three considered distributions. The left panel corresponds to the case when  $f$ is constant; the middle panel - $f$ is from uniform distribution; right panel - $f$ is from normal distribution. Bottom panels: the estimated values of fault frequencies for three analyzed distributions.}
\label{simul_line_plot}
\end{figure}

In Fig. \ref{simul_line_plot} in the upper panel, we present theoretical frequency values for the $100$ segments. The constant value of $f$ is $30$ Hz for all segments. In the second case, $f$ was generated from a uniform distribution in the range $[29,31]$ independently for each segment. In the third case, $f$ was generated from the normal distribution with $\mu=30$ and $\sigma=0.33$ for each segment independently. In the lower panel, we can see the estimated values of $f$ for 100 segments, 5 s each. The most important thing is that even if we have constant frequencies for all of the segments, as a result of the estimation, we can get varying values. For uniform and normal distributions, the estimated values are very close to the theoretical ones, but they are not the same. In these cases (uniform and normal distribution of $f$), we have relatively large variances, so identification is easy. For smaller variances, we can have a worse classification accuracy.

By analyzing only the estimated values of fault frequency, it is difficult to identify if they correspond to the uniform or normal distribution. 
To clearly confirm the corresponding distribution, we can compare the theoretical PDF with KDE calculated based on the estimated values; see Fig. \ref{simul_density}.

\begin{figure}[H]
\centering
    \includegraphics[width = 15cm]{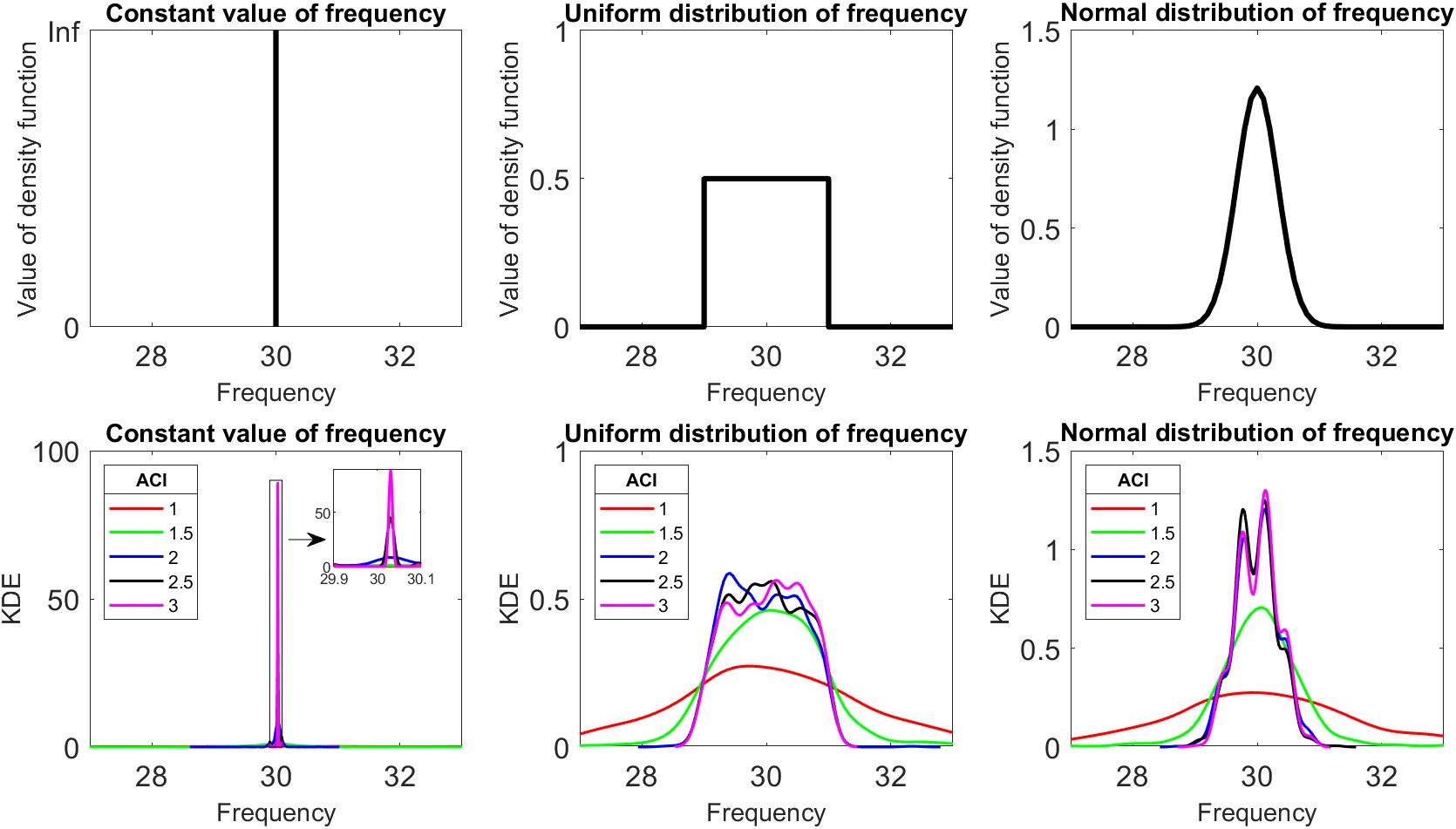}
\caption{Comparison of PDF and KDE for different frequency distributions and different ACI}
\label{simul_density}
\end{figure}

Another very important parameter is the ACI. In Fig. \ref{simul_density} in the upper panel, we present PDF for particular distributions of $f$ and in the lower panel, we compare KDE for different distributions of $f$ and different ACI. {The ACI} has a direct impact on SNR, because the amplitude of the noise is fixed.  It is important to note that the maximum value of PDF for a constant frequency is infinite. This means that the whole probability is concentrated at one point.

In the left panel (constant value of $f$), even for a large ACI (when it is easy to detect the peaks), we get different estimated values of $f$, where the expected value is $30Hz$. The variance {of the frequency values} is very low, but is still not equal to $0$. For lower ACI, we can see that the variance increased significantly. We have a similar situation for KDE for uniform and normal distributions. For large ACI, they are close to the theoretical PDFs, but the difference is especially visible for ACI $=1$. In that case, both considered distribution plots look similar. Because for low ACI, peak detection is difficult, the true distribution of $f$ therefore does not influence the estimated values.

{In Fig. \ref{simul_MSE}. we can see the impact of ACI and segment length on estimation accuracy. To measure this impact, we use MSE.} 

\begin{figure}[H]
\centering
  \includegraphics[width = 15cm]{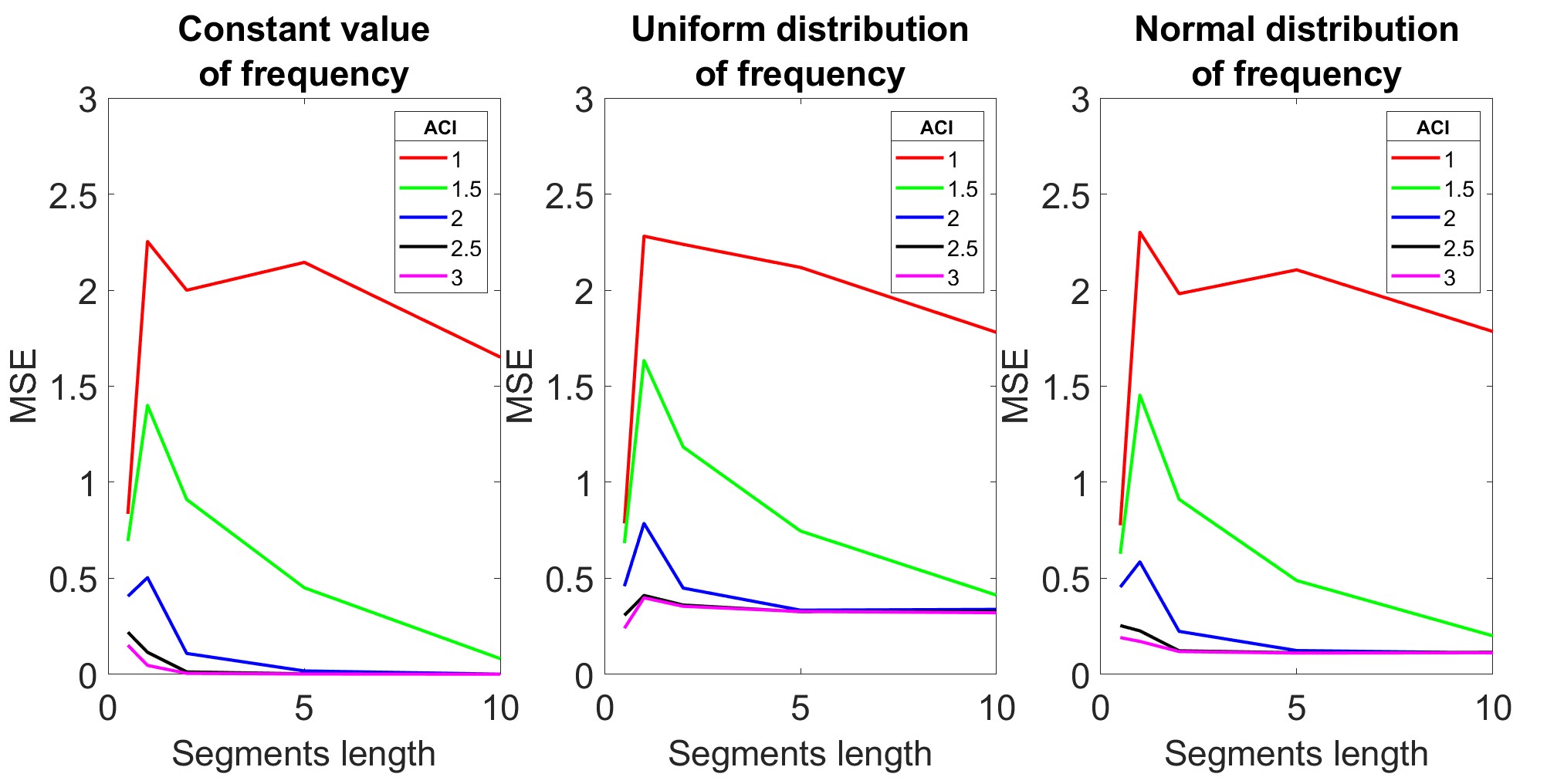}
\caption{Comparison of MSE for different distributions of frequency, different ACI and different segment length}
\label{simul_MSE}
\end{figure}

If we have a very low ACI, the theoretical distribution does not influence the estimated value. In every plot in Fig. \ref{simul_MSE}, the MSE plots for ACI = 1 look very similar. The second observation is that if we have longer segments length, then we have also lower MSE. For a constant value of $f$, MSE converges to $0$ for increasing segment length (especially for relatively large ACI). On the other hand, MSE does not converge to $0$ for both normal and uniform distributions of $f$. The reason is that we calculate the error as the difference between observed value and expected value of the distribution; thus, even if we perfectly estimated $f$, we will get a non-zero error (depending on randomly generated true value of the frequency).

The next step in our procedure is to check if the frequency of cyclic impulsive signals is constant. As shown in Fig. \ref{block_diagram}, we have to check whether the variance is lower than the threshold, and if it is not, then we perform the chi-squared test for the equality of variances. To show the precision of our methodology, we generate $10000$ independent signals for each combination of ACI = $\{1,1.5,2,2.5,3\}$ and segment lengths = $\{0.5s,1s,2s,5s,10s\}$. Then we divide the mentioned simulated signals into $100$ parts consisting of $100$ signals each. Next, we estimate frequencies and calculate variances of estimated values. Finally, because we had $100$ parts, we were able to check how many times the frequencies were incorrectly classified as a constant value. Tables \ref{Percentage_unif} and \ref{Percentage_norm} show the results for the uniform and normal distribution of the frequency, respectively. 

\begin{table}[H]
    \centering
    \begin{tabular}{|c|c|c|c|c|c|}
    \hline
        & \multicolumn{5}{c|}{Segments length} \\ \hline
        \multicolumn{1}{|c|}{ACI} & 0.5 s &  1 s & 2 s & 5 s & 10 s \\ \hline 
        1.0 & 100 & 93 & 98 & 94 & 80 \\ \hline
        1.5 & 99 & 75 & 60 & 5 & 0 \\ \hline
        2.0 & 54 & 10 & 0 & 0 & 0 \\ \hline
        2.5 & 1 & 0 & 0 & 0 & 0 \\ \hline
        3.0 & 0 & 0 & 0 & 0 & 0 \\ \hline
    \end{tabular}
             \caption{Percentage of wrongly classified distributions of frequency as constant value, generated from uniform distribution (U(29,31)), based on 100 signals with 100 segments each}
             \label{Percentage_unif}
\end{table}

Table \ref{Percentage_unif} shows the classification results (constant frequency or not) for uniform distribution. We can see here the same situation as in Fig. \ref{simul_MSE} -- if we do not have a high enough ACI (e.g. ACI $= 1$), even for long segments, we are not able to distinguish constant frequencies from randomly generated frequencies from uniform distribution. A slightly different situation is for ACI $= 1.5$. In that case, for segments of 5 and 10 s, the classification accuracy is very high. Unfortunately, similarly for shorter segments, our method failed. On the other hand, if we have a very high ACI, for all segments considered in this paper, we get the correct classification.

\begin{table}[H]
    \centering
    \begin{tabular}{|c|c|c|c|c|c|}
    \hline
        & \multicolumn{5}{c|}{Segments length} \\ \hline
        \multicolumn{1}{|c|}{ACI} & 0.5 s &  1 s & 2 s & 5 s & 10 s \\ \hline 
        1.0 & 100 & 91 & 96 & 94 & 95 \\ \hline
        1.5 & 100 & 94 & 87 & 63 & 0 \\ \hline
        2.0 & 93 & 63 & 0 & 0 & 0 \\ \hline
        2.5 & 63 & 0 & 0 & 0 & 0 \\ \hline
        3.0 & 4 & 0 & 0 & 0 & 0 \\ \hline
    \end{tabular}
             \caption{Percentage of wrongly classified distributions of frequency as constant value, generated from normal distribution (N(30,$0.33^2$)), based on 100 signals with 100 segments each}
              \label{Percentage_norm}
\end{table}
    
Very similar results as for uniform distribution, we have for normally distributed frequencies (see Table \ref{Percentage_norm}). We are still not able to recognize a normal distribution if we have a very small ACI, but with increasing ACI and segment length, the percentage of wrongly classified distributions is decreasing. For ACI $= 3$, only for $0.5$s segments, we were unable to perfectly classify the frequency distributions, but the accuracy was still very high ($96\%$).

At the end of this section we present Table \ref{Thresholds_table} with thresholds for different segment lengths and different ACI calculated based on the simulated data (see the procedure presented in Fig. \ref{Threshold block diagram} ). We recall that the threshold is used in the procedure for testing the variation of fault frequency presented in the previous section. The derived thresholds can be useful for the analysis of real data.
\begin{table}[H]
    \centering
    \begin{tabular}{|c|c|c|c|c|c|}
    \hline
        & \multicolumn{5}{c|}{Segments length} \\ \hline
        \multicolumn{1}{|c|}{ACI} & 0.5 s &  1 s & 2 s & 5 s & 10 s \\ \hline 
        1 & 0.5842 & 2.0353 & 1.9916 & 2.1184 & 1.6484 \\ \hline
        1.5 & 0.4944 & 1.3307 & 0.9052 & 0.4505 & 0.0809 \\ \hline
        2 & 0.3030 & 0.4980 & 0.1082 & 0.0163 & 0.0003 \\ \hline
        2.5 & 0.1587 & 0.1129 & 0.0118 & 0.0009 & 0.0000 \\ \hline
        3 & 0.0888 & 0.0411 & 0.0037 & 0.0003 & 0.0000 \\ \hline
    \end{tabular}
             \caption{Thresholds calculated based on simulated signals of different lengths and different ACI.  }
             \label{Thresholds_table}
\end{table}

\section{Real signals analysis}\label{real}
In this section we present the analysis for several real signals with potential different distributions of background noise and different nature of speed variation.



\subsection{Vibration signal (Vib2) from a test rig}
The first data set was acquired from a laboratory test rig (see Fig. \ref {rig}). The experiments assumed linear growth of speed (forced in manual way by speed controller), stationary regime (with constant speed), and finally the speed was linearly decreased up to stopping the machine.

\begin{figure}[H]
\centering
  \includegraphics[width=10 cm]{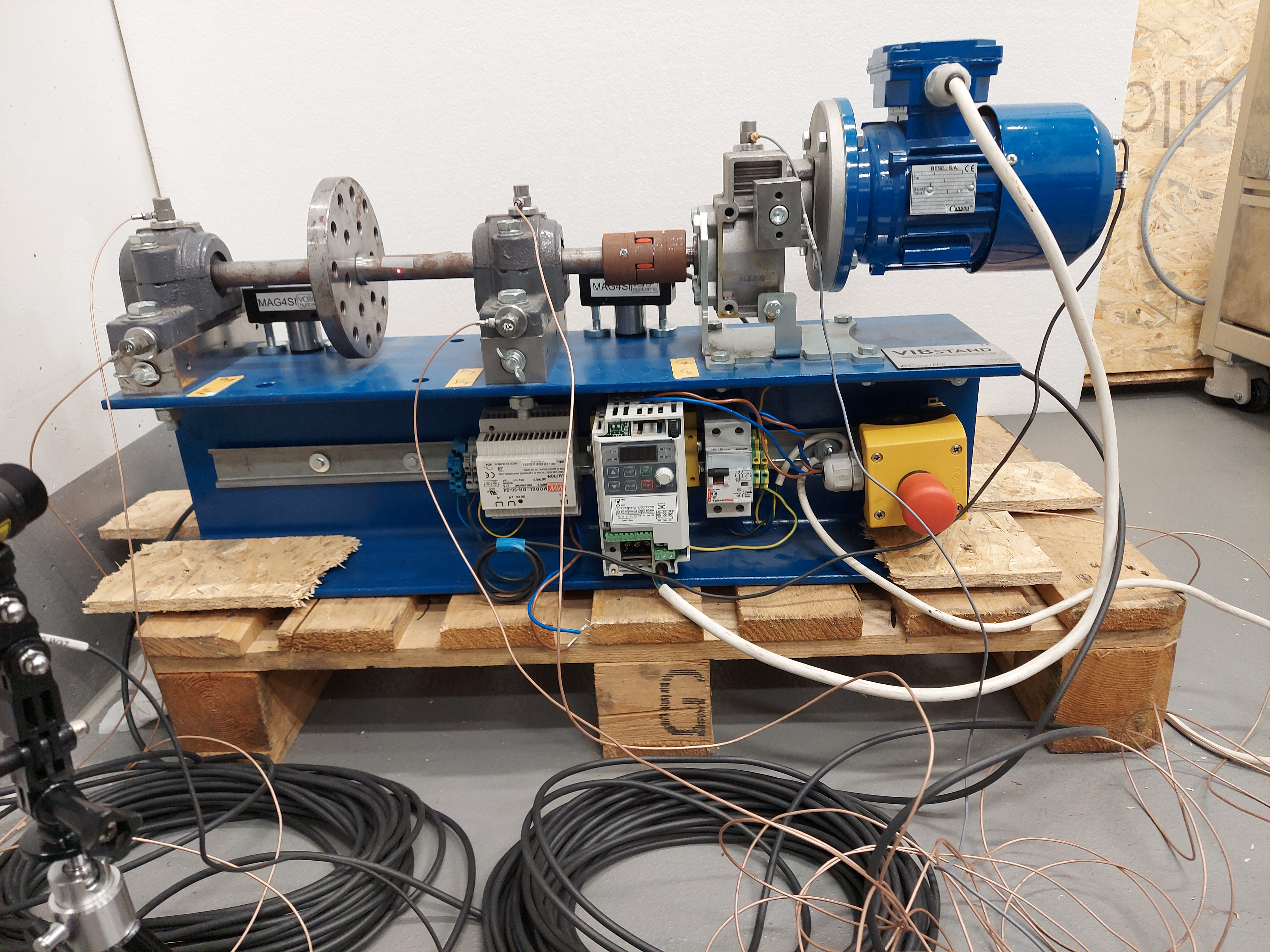}
    \caption{Test rig examined in the experiment.}
    \label{rig}
\end{figure}

The signal called Vib2 is presented in Fig. \ref{vib2_whole}.

\begin{figure}[H]
\centering
    \includegraphics[width = 10cm]{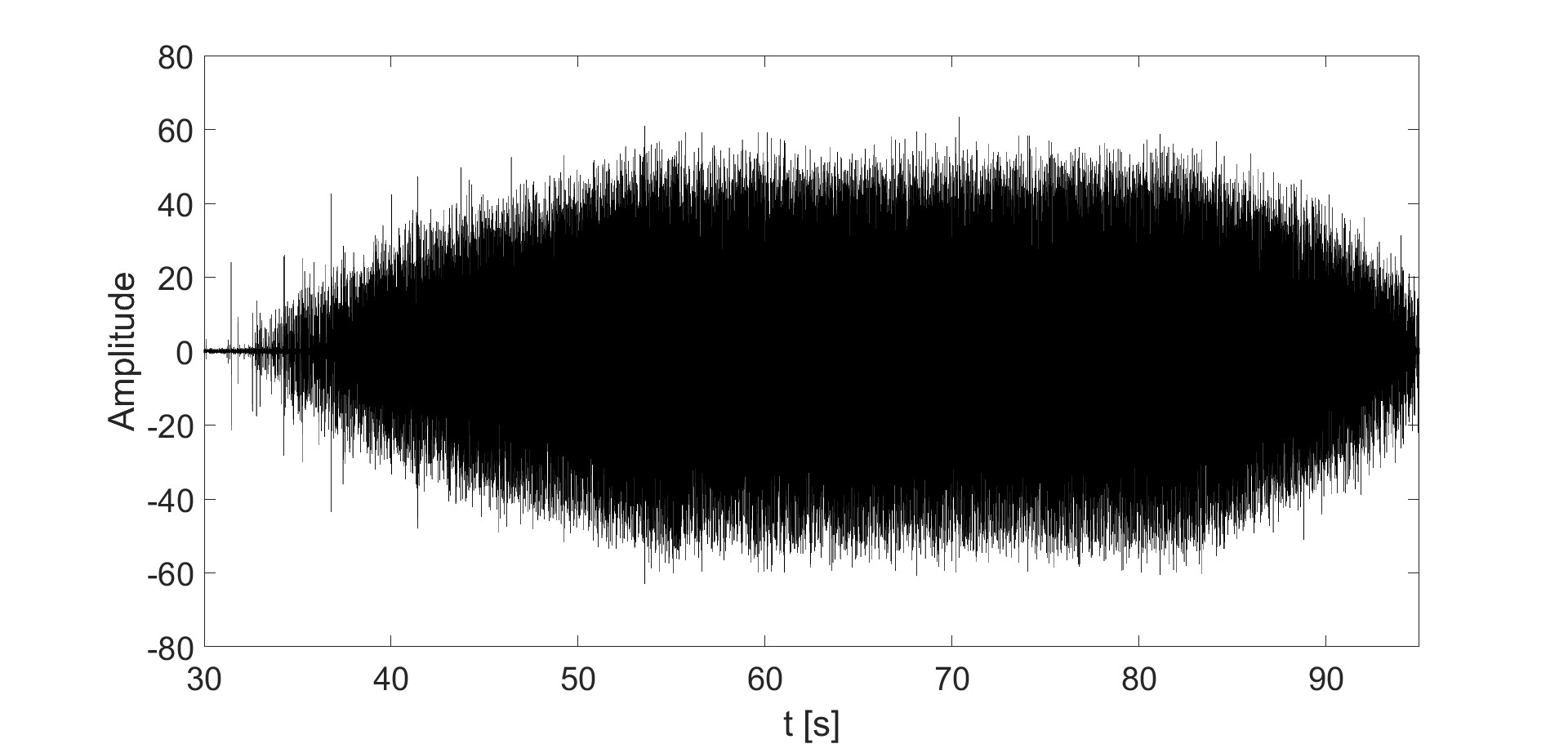}
    \caption{Analysed part of Vib2\textcolor{red}{ }}
    \label{vib2_whole}
\end{figure}

For further analysis, the signal has been divided into three parts: the first part covers an increasing speed regime (33s -- 53s), the middle part when the machine works with constant speed (67s -- 83s), and the third part is related to decreasing the speed (83s -- 96s). 


\begin{figure}[H]
\centering
    \includegraphics[width = 10cm]{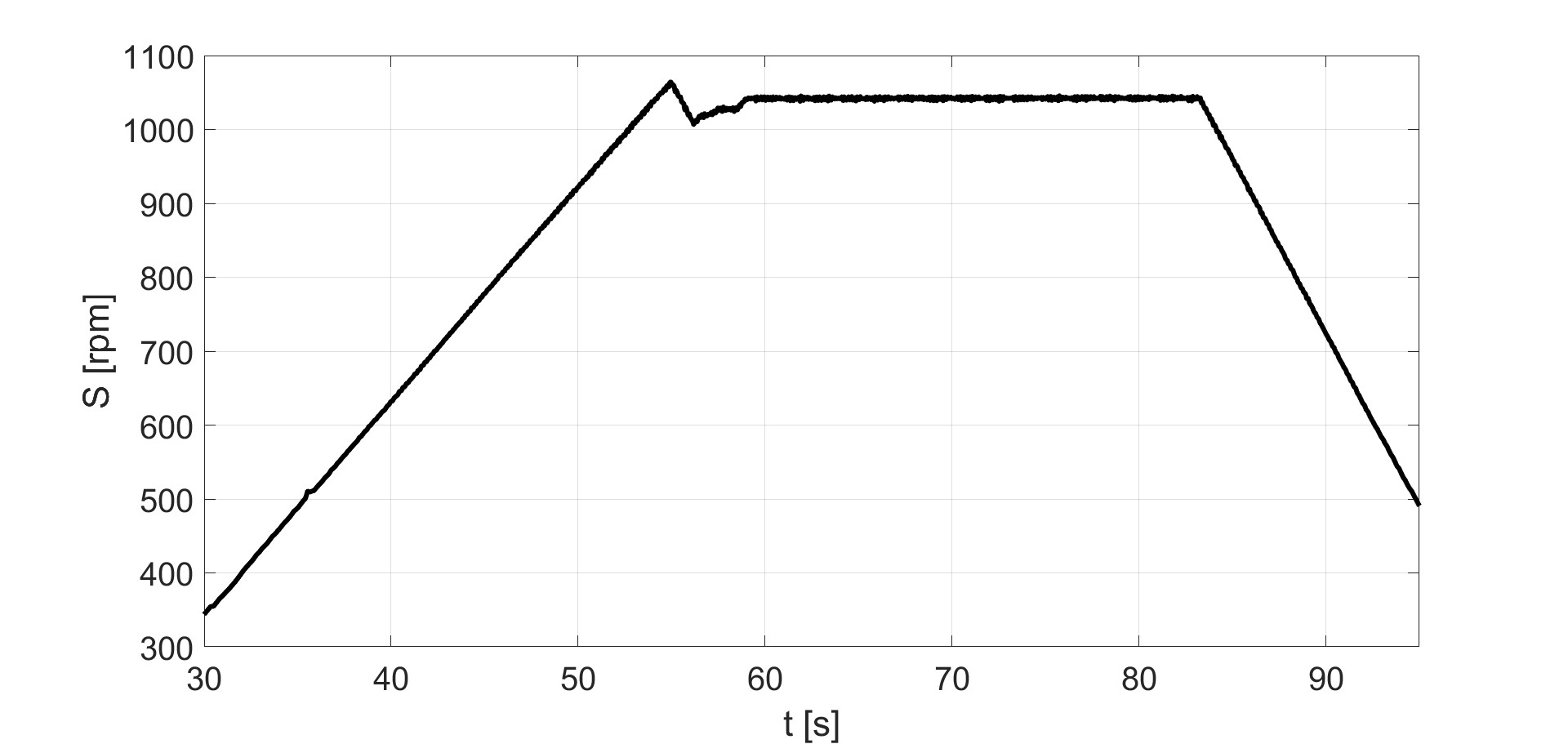}
    \caption{Speed profile for Vib2 }
    \label{vib2_speed_profile}
\end{figure}

As we can see in Fig. \ref{vib2_speed_profile} during the first part of the signal, the speed was, in fact, linearly increasing from approximately $350$ to about $1050$ [rpm]. In the middle part, it is approximately constant and equals $\sim 1050$ (note the change of speed profile plot between 53 and 67 s -- it will not be used for analysis). The third part of the signal starts at 83 s up to 96 s, and the speed decreases there from $1050$ to $500$.

\begin{figure}[H]
\centering
    \includegraphics[width = 12cm]{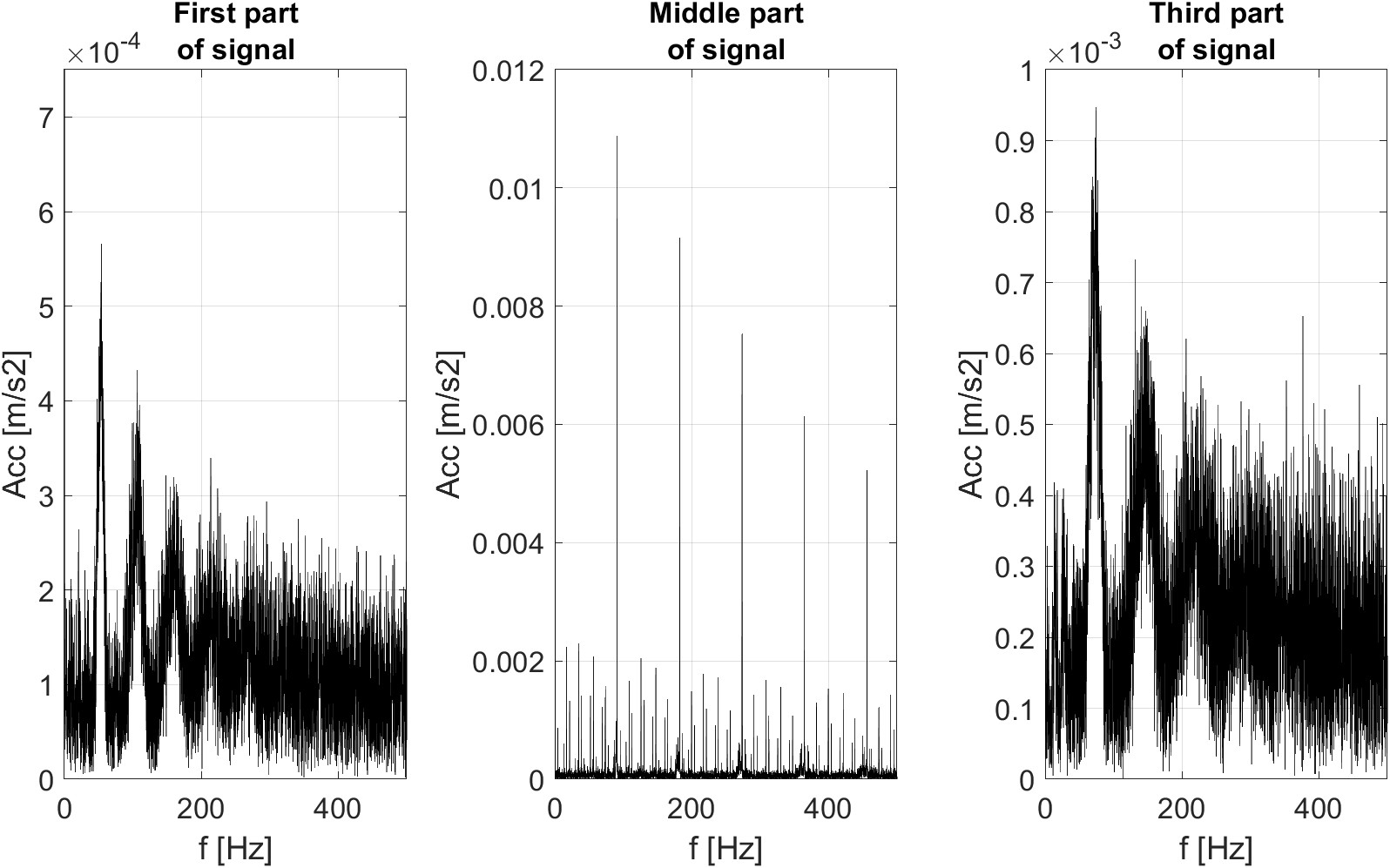}
    \caption{Envelope spectrum for first part of signal, middle part of signal (when machine works with constant speed) and for third part of signal, respectively}
    \label{env_Vib2}
\end{figure}

In Fig. \ref{env_Vib2}, plots of the envelope spectra are shown for the first, middle, and third parts of the signal, respectively. In the first and third part, we have a very similar situation. This is because in the first case the speed and the amplitude are increasing, and in the third case they are decreasing, but the behavior is almost the same. The most important information that we can see here is that the spectra of the signals are very blurry. That means that detecting peaks properly is very unlikely. We can estimate the frequency on the basis of those plots, but the error could be high. In these cases only the first three peaks are visible, but the next peaks are covered by noise. On the other hand, the plot for the middle part gives very clear results. The amplitude of peaks is much greater than the amplitude of noise. Exactly opposite to the first and third parts of the signal, every peak in the graph is visible. On the basis of these data, we are able to estimate the frequency very precisely.

Now, we can move on to analysis of all three parts separately. For Vib2 data, we consider $0.5$ s, $1$ s and $2$ s lengths of segments. Thus, we are dividing the first, second and third part of the signal into shorter segments (assuming local stationarity of the speed) and estimate the frequency for each of them. In Fig. \ref{vib2_line_plot_first_part} we can see how the estimated values change over time for the first part of the signal for the mentioned lengths of the segments.

\begin{figure}[H]
\centering
    \includegraphics[width = 12cm]{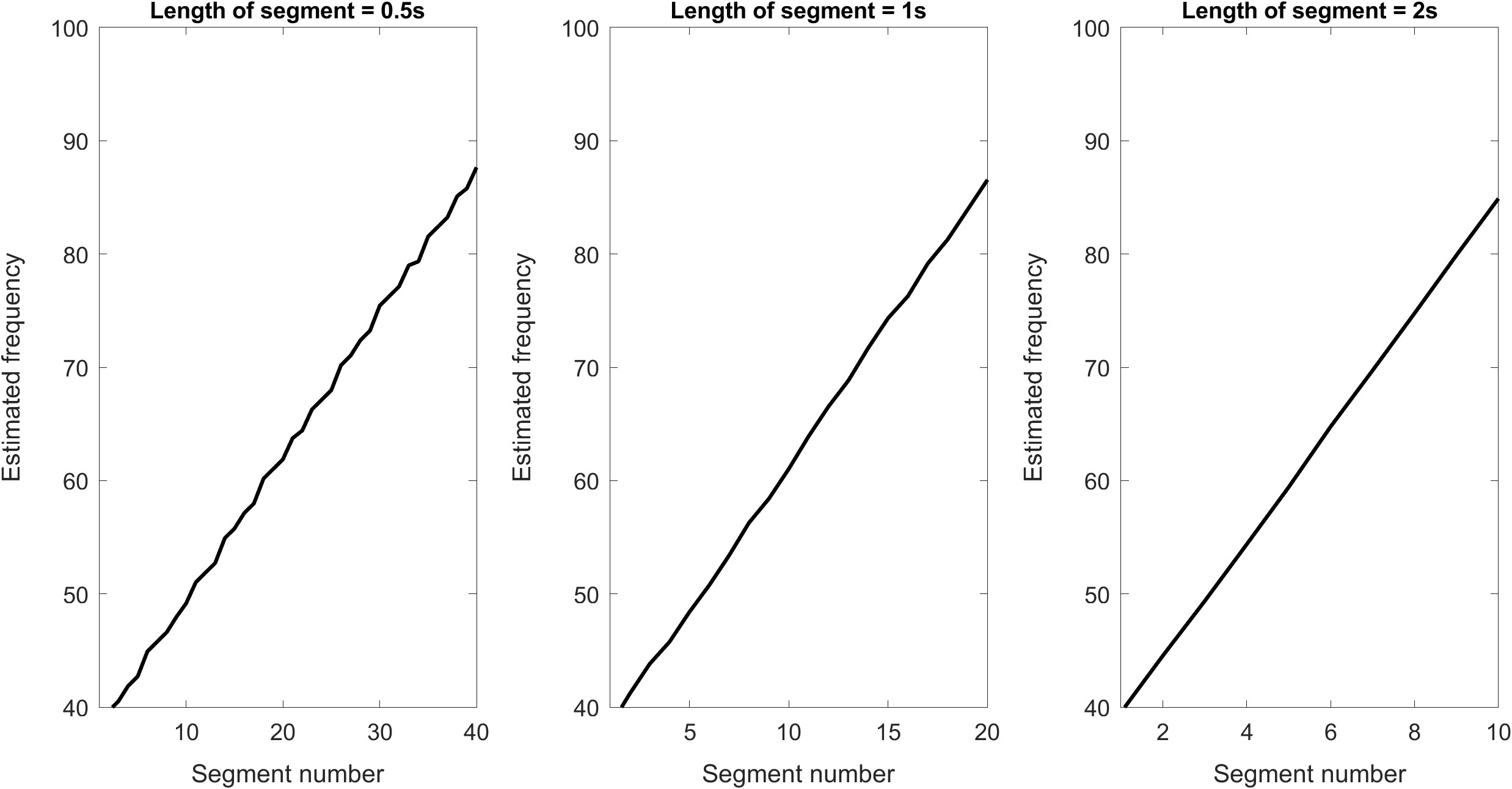}
    \caption{First part of signal}
    \label{vib2_line_plot_first_part}
\end{figure}

In Fig. \ref{vib2_line_plot_first_part}, we can see that for each consecutive segment, the estimated value is greater than for the previous segments. All three plots look very similar. For 0.5s segments, there are small deviations from the linear function, but for 1s and 2s segments, it is much closer. At the beginning of the signal, the frequency was estimated as 40Hz and finally achieved value about 87Hz.

Next, we can repeat the procedure for the middle part of the signal. We consider the same segment length as before. The estimated frequency values for that part of the signal are shown in Fig. \ref{vib2_line_plot_middle_part}.

\begin{figure}[H]
\centering
    \includegraphics[width = 12cm]{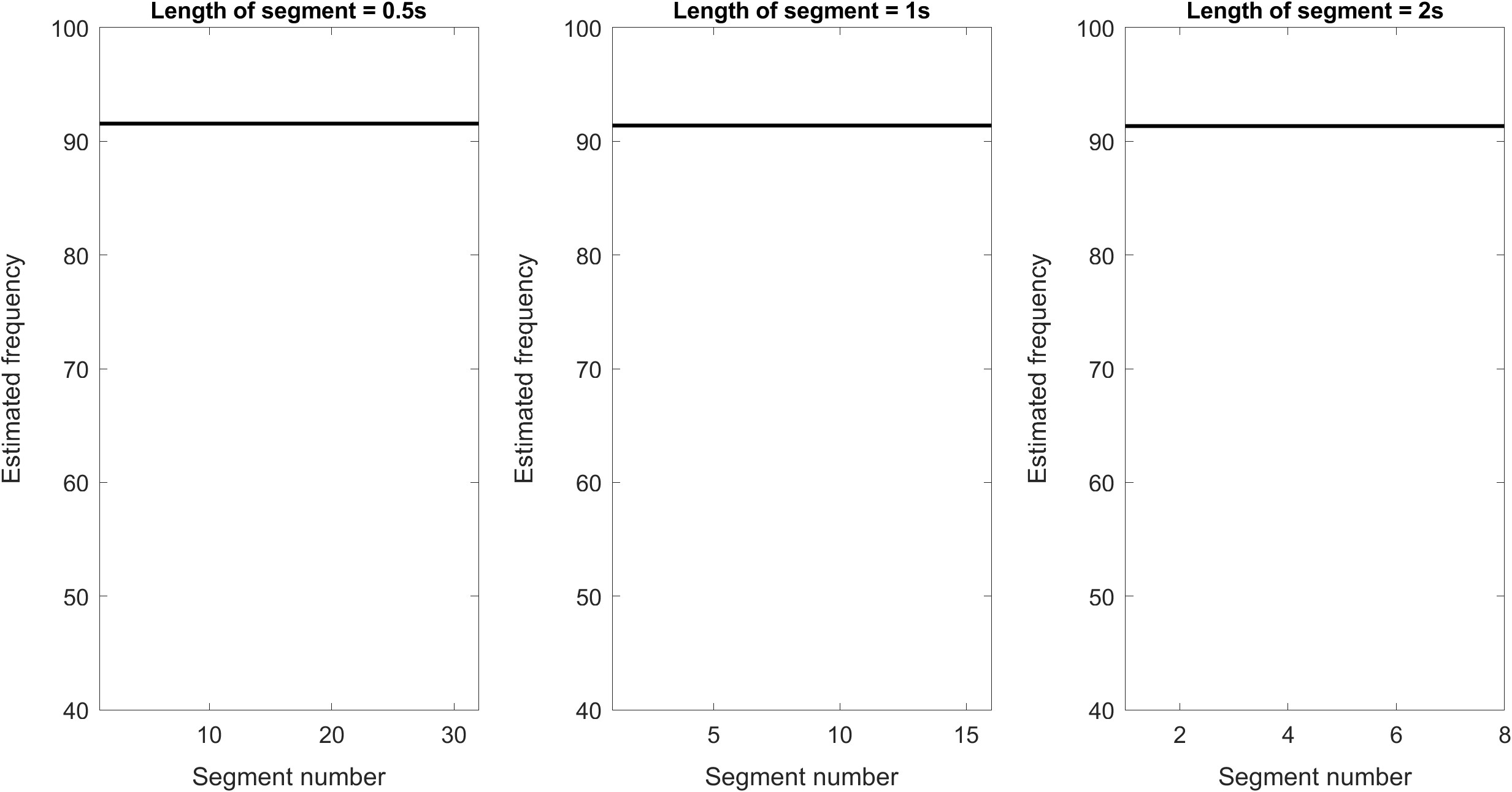}
    \caption{Middle part of signal}
    \label{vib2_line_plot_middle_part}
\end{figure}

For all plots, we have obtained the same results. Consecutive values form a special case of a linear function, the constant function. For each segment, the estimated values are the same and equal to about 92Hz. There are small differences for different lengths of segments.

For the third part of the signal, we have the reverse situation than for the first part (see Fig. \ref{vib2_line_plot_third_part}). It is not surprising if we look at the speed profile of Vib2. The behavior of the estimated $f$ for that part of the signal is easy to predict based on the first part.{ Estimated values decrease from 90 to 40.}

\begin{figure}[H]
\centering
    \includegraphics[width = 12cm]{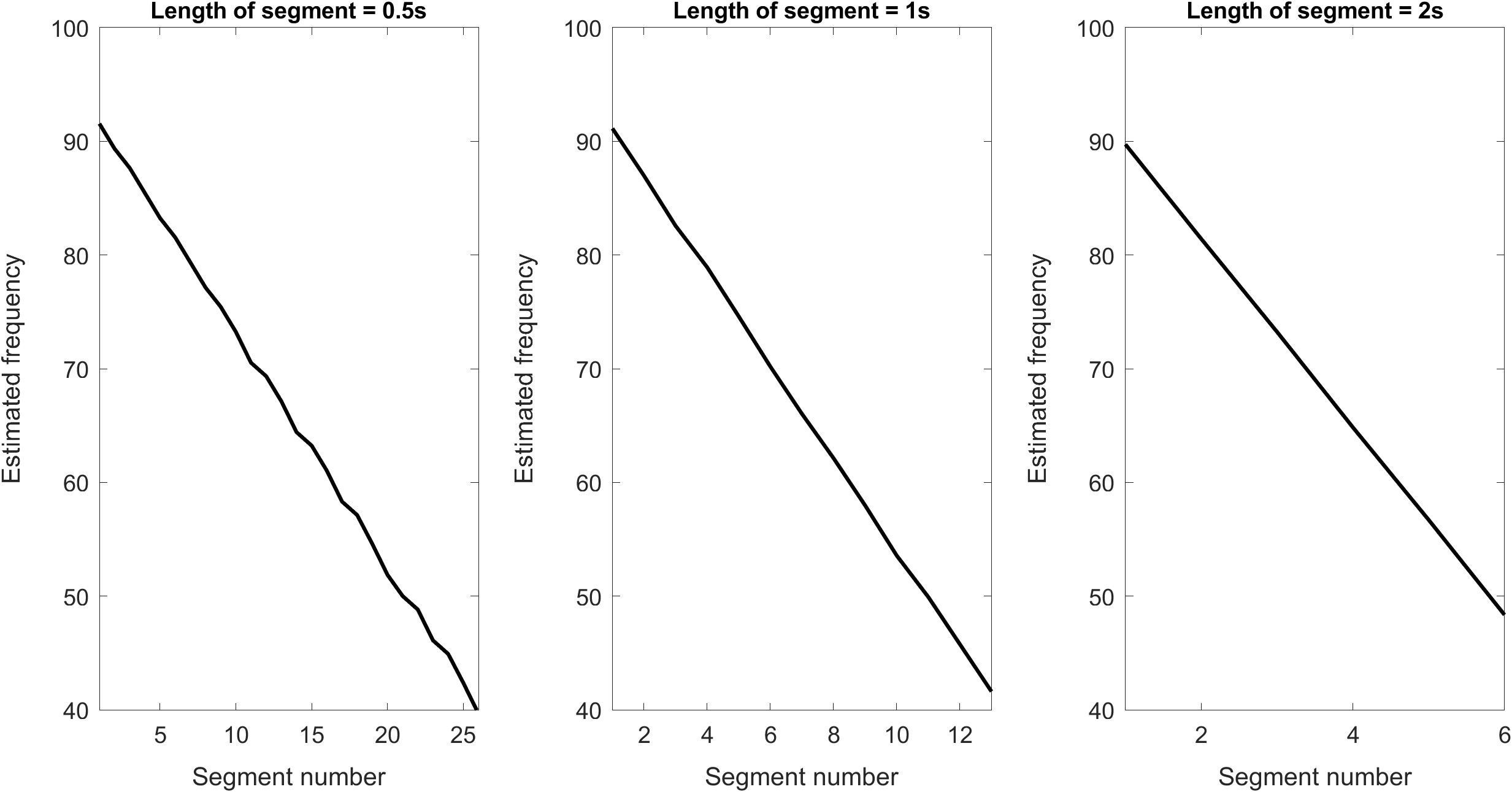}
    \caption{Third part of signal}
    \label{vib2_line_plot_third_part}
\end{figure}

Only for 2s-segments, the lowest estimated value is closer to 50 than to 40, because there is included longer part of signal than for 0.5s and 1s segments. Thus, we cannot detect rapidly changing values of $f$. The results of the estimation opposite to the first part of the signal decrease over time. 

Thanks to considering estimated frequency (for non-overlapped segments) as time series, on simple visualization, we can see the general behavior of signal. Unfortunately, it could be very hard to recognize if the distribution of $f$ is closer to normal or uniform. Sometimes it could be even hard to decide whether frequency could be considered as constant but due to errors connected with analysis and estimation, variance is different than 0 or it is not constant. For this purpose, one of the most common ways is to perform a statistical test based on a PDF or cumulative density function. Graphs of the first one are shown in Figs. \ref{vib2_density_plot_first_part}, \ref{vib2_density_plot_middle_part} and \ref{vib2_density_plot_third_part} for the 1st, 2nd and 3rd considered parts of the signal, respectively.

\begin{figure}[H]
\centering
    \includegraphics[width = 12cm]{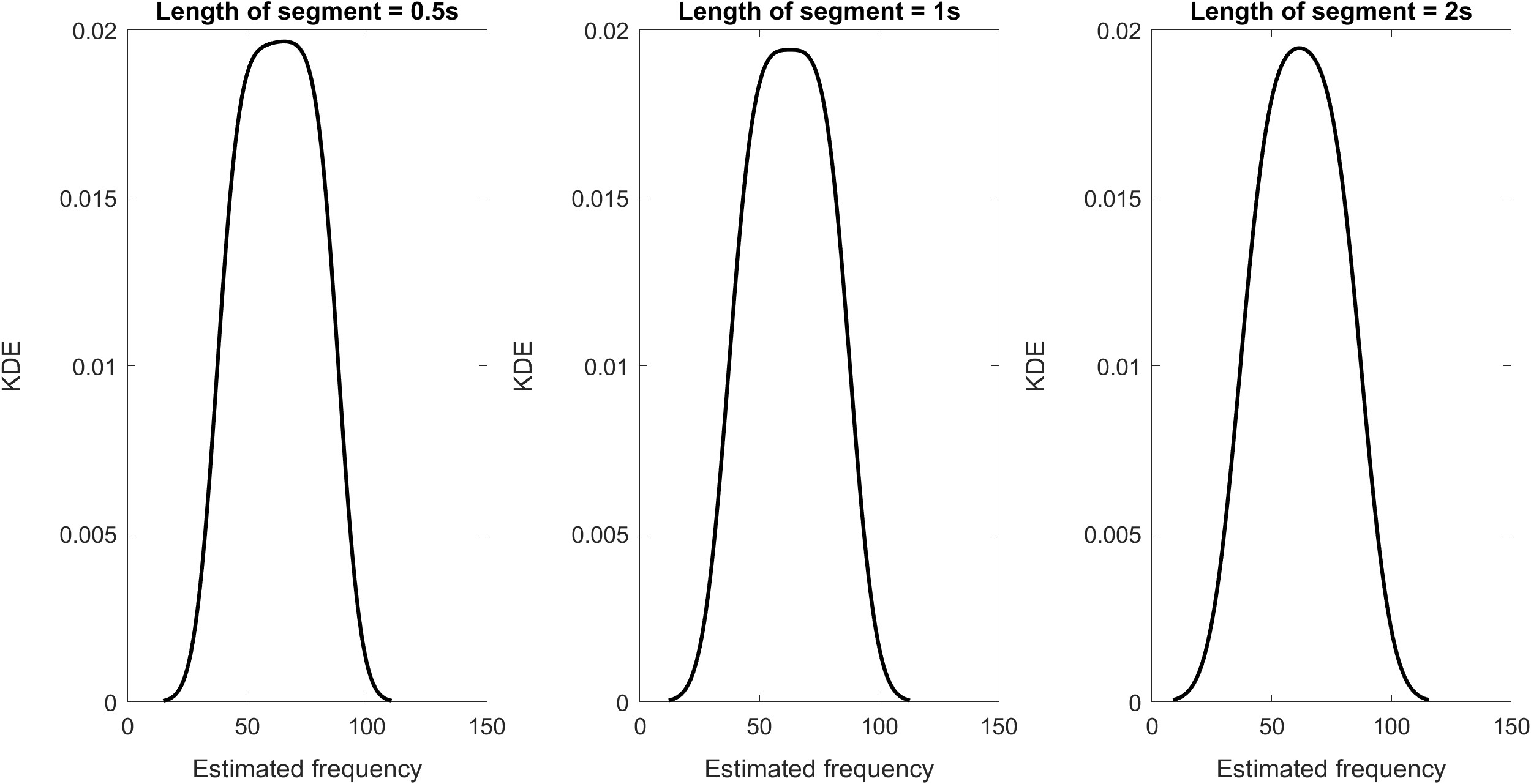}
    \caption{Comparison of KDE for the first part of signal}
    \label{vib2_density_plot_first_part}
\end{figure}

Fig. \ref{vib2_density_plot_first_part} shows KDE for the first part of the signal. Note that in many cases input speed measurement is not available, so we do not know if speed is varying or not. Moreover, with minor speed fluctuation (or even almost constant speed) after a dozen measurements, one will obtain a distribution of frequencies rather than just one exact value of the fault frequency.

We can definitely say that it cannot be denoted as constant $f$ (in the next part of the paper we will prove that). Based on a time series of estimated values, we could conclude that KDE will look like a uniform distribution. On the other hand, based on the plots in Fig. \ref{vib2_density_plot_first_part}, KDE looks like a combination of normal and uniform distributions. 
The important fact is that the values strictly increase in time (because we manually increased the speed). 
Therefore, this is one of the main reasons why the frequency is changing. This shows that if we have varying speed, we can get similar results (in the context of shape of KDE) like for constant speed, but the variance is much higher. 

This is the reason we apply the statistical test to confirm this hypothesis.

\begin{figure}[H]
\centering
    \includegraphics[width = 12cm]{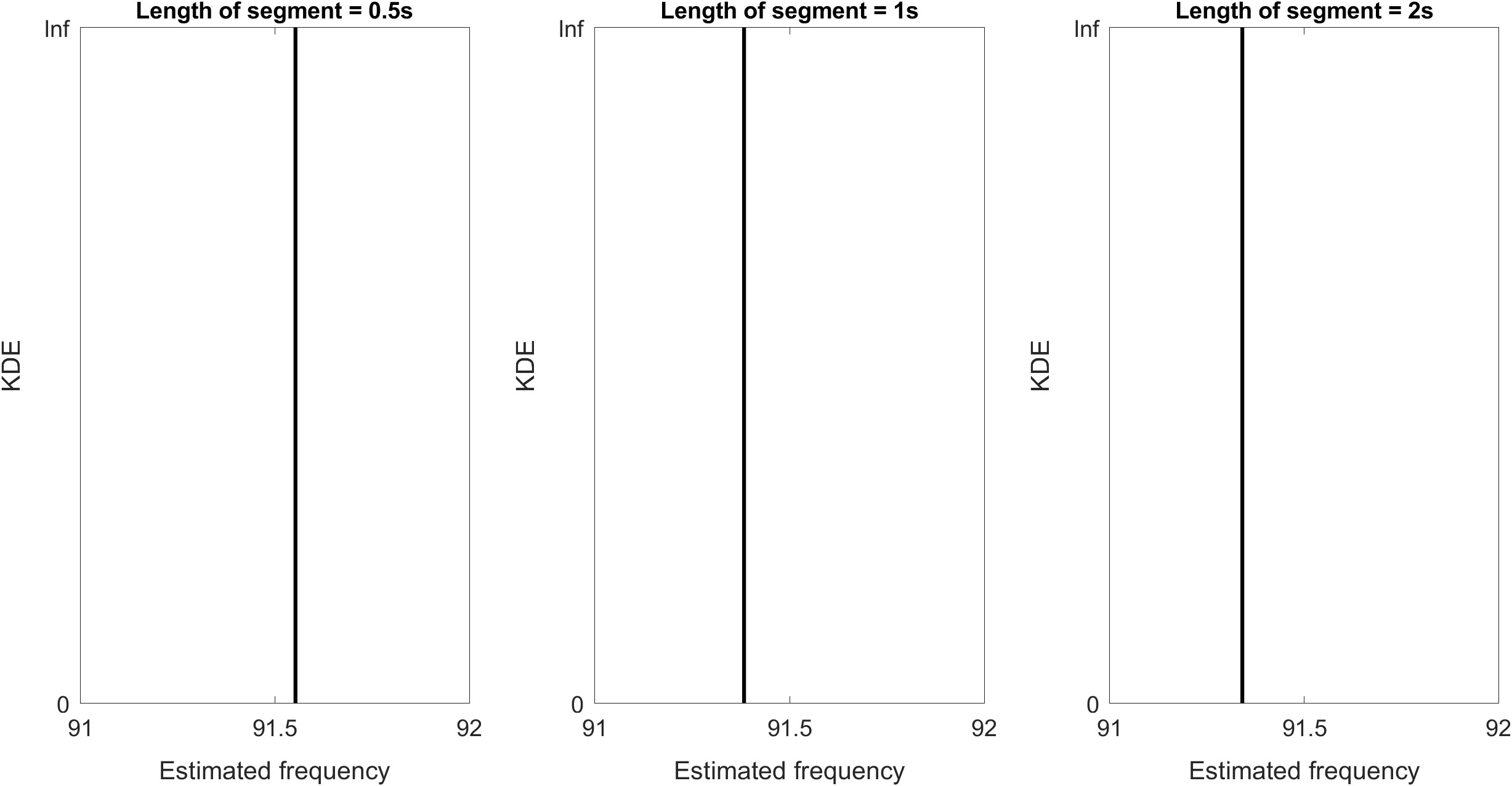}
    \caption{Comparison of KDE for the middle part of signal (when machine works stably)}
    \label{vib2_density_plot_middle_part}
\end{figure}

The opposite situation is for the middle part of the signal. This is because the machine worked with constant speed. We can see in Fig. \ref{vib2_density_plot_middle_part} that for all of the considered length of segments, for every segment, we get the same value. For 0.5s it is above 91.5Hz and for 1s, 2s below that value but the difference between them is very small. Note that the maximum values of KDE are infinite. That means that we get the same estimated value of $f$ for every segment. 

Lastly, in Fig. \ref{vib2_density_plot_third_part} KDE is shown for the third part of the signal. 

\begin{figure}[H]
\centering
    \includegraphics[width = 12cm]{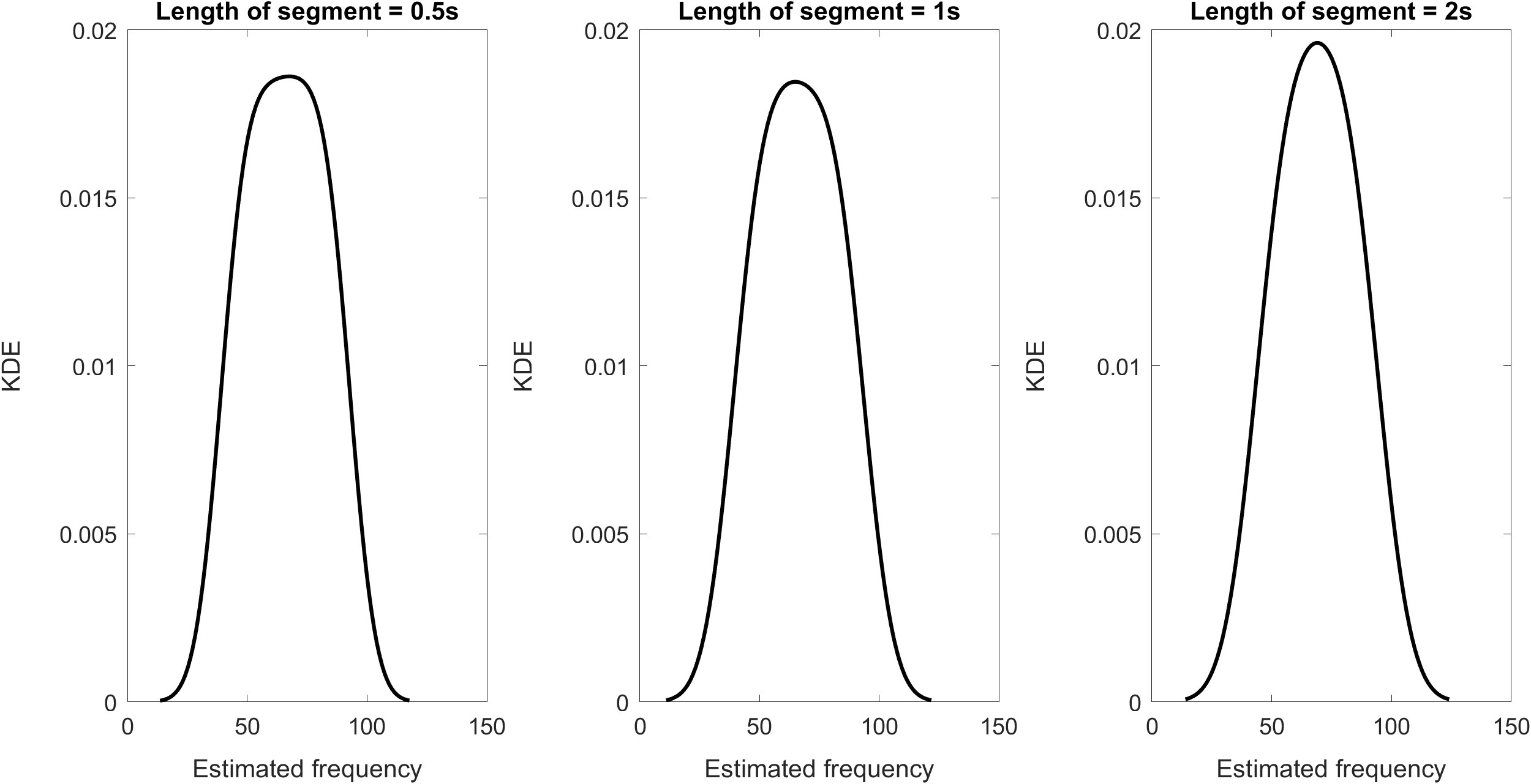}
    \caption{Comparison of KDE for the third part of signal}
    \label{vib2_density_plot_third_part}
\end{figure}

During analysis of line plots for consecutive segments, we mentioned that for the third part, we have reverse behavior of estimated frequencies -- values decreasing in time. However, the order of the values does not influence KDE. That means that we can expect very similar plots for the third part as for the first part. Moreover, for all of the considered lengths of segments, the graphs are almost the same with only small differences. 


In this subsection, we were analyzing three parts of Vib2. All have a specific behavior. The first part is related to the start up regime, and the last one was related to machine shutdown. The middle part describes the machine working under constant operating conditions.
In the next subsection, we will repeat the procedure for other types of signal (for machine working in non-stationary operations with speed/load fluctuation). 

\subsection{Industrial vibration signals with local damage}

In this subsection we analyze two real data sets from industrial machine (two stage gearbox used in drive unit).
{Firstly, we present speed profiles and time series, see Figs. \ref{profile_reale_analizy}, \ref{Sig1}, \ref{Sig2}}. This can show us what are the characteristics of considered signals. Second, we perform segmentation and estimate frequency separately for each segment. To visualize how the estimated $f$ changes over time, the results will be shown as time series. Finally, we calculate KDE to check the distributions of frequencies.


\begin{figure}[H]
\centering
    \includegraphics[width = 10cm]{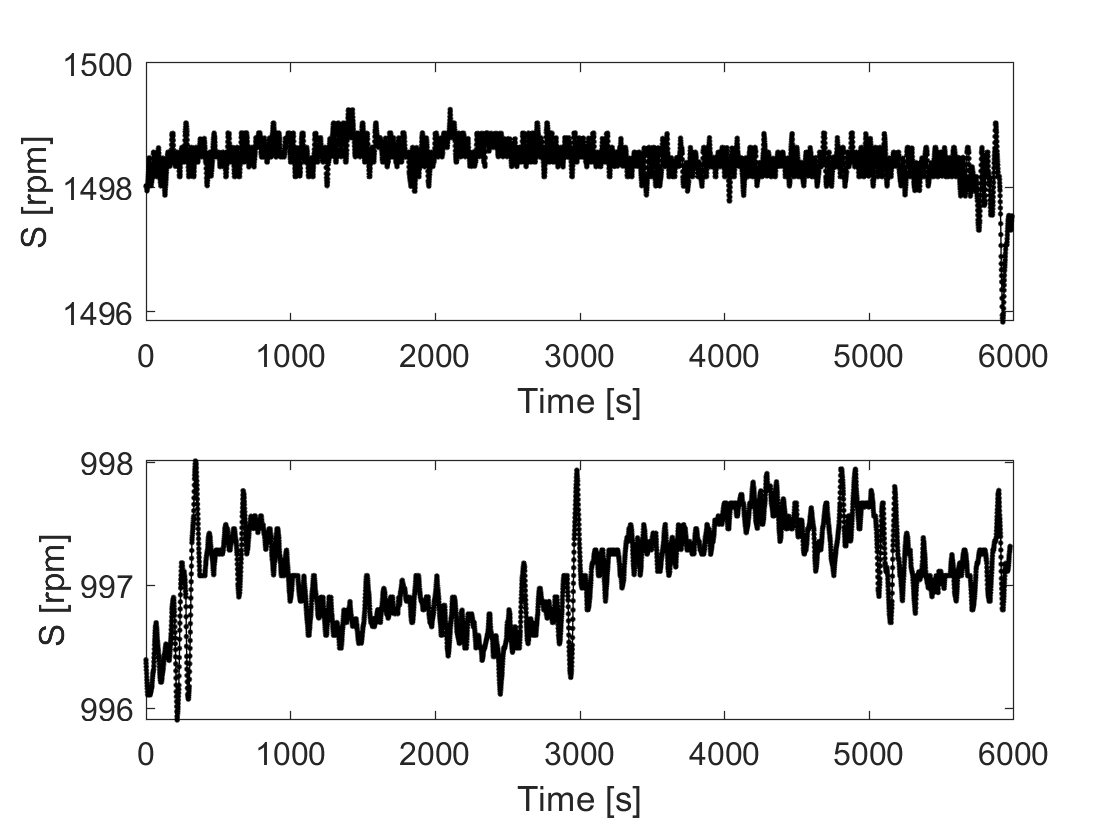}
    \caption{Two exemplary speed profiles used for analysis: top: nearly constant speed, bottom -- variation of the speed }
    \label{profile_reale_analizy}
\end{figure}


\begin{figure}[H]
\centering
    \includegraphics[width = 10cm]{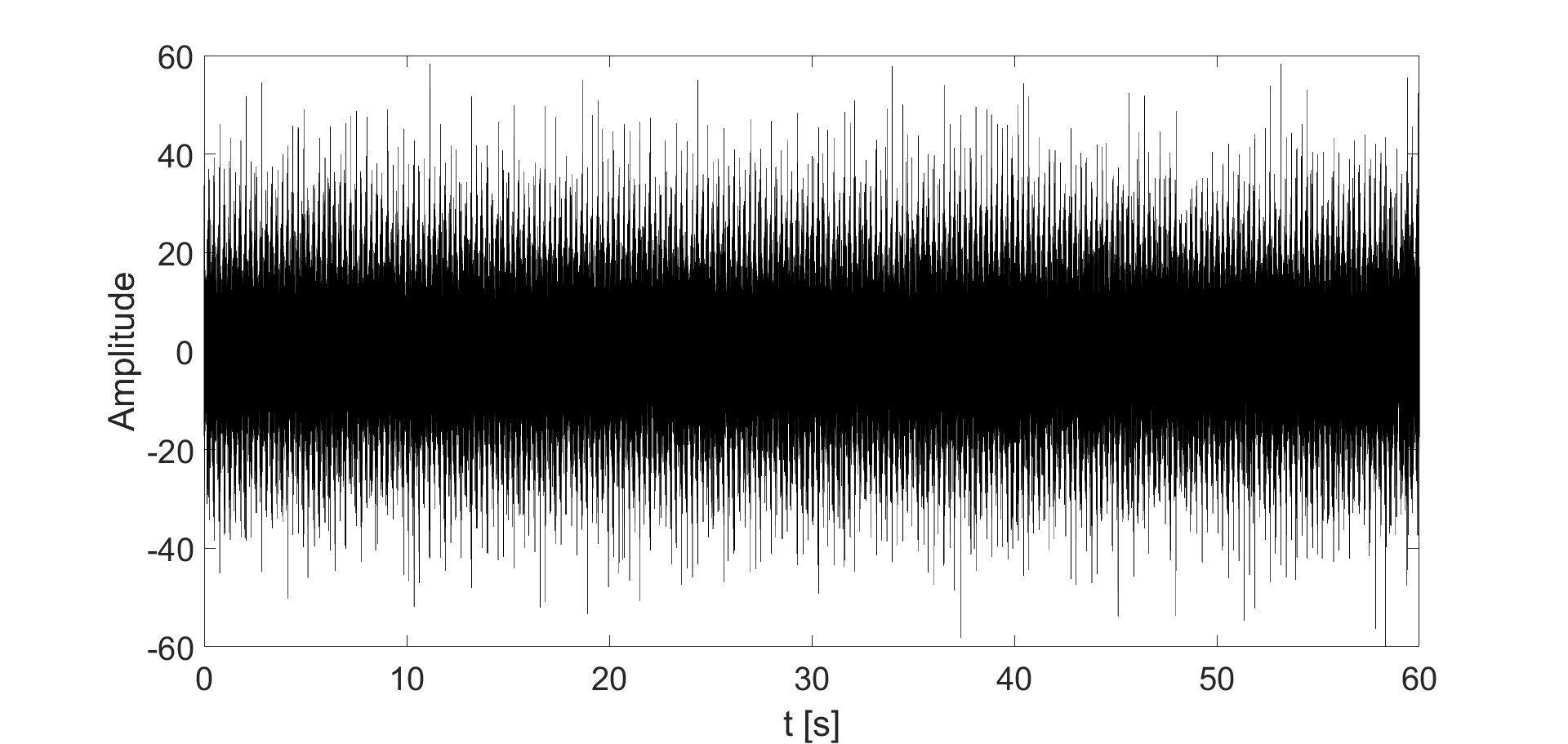}
    \caption{First example of real vibration signal with minor speed fluctuation (Sig1)}
    \label{Sig1}
\end{figure}

\begin{figure}[H]
\centering
    \includegraphics[width = 10cm]{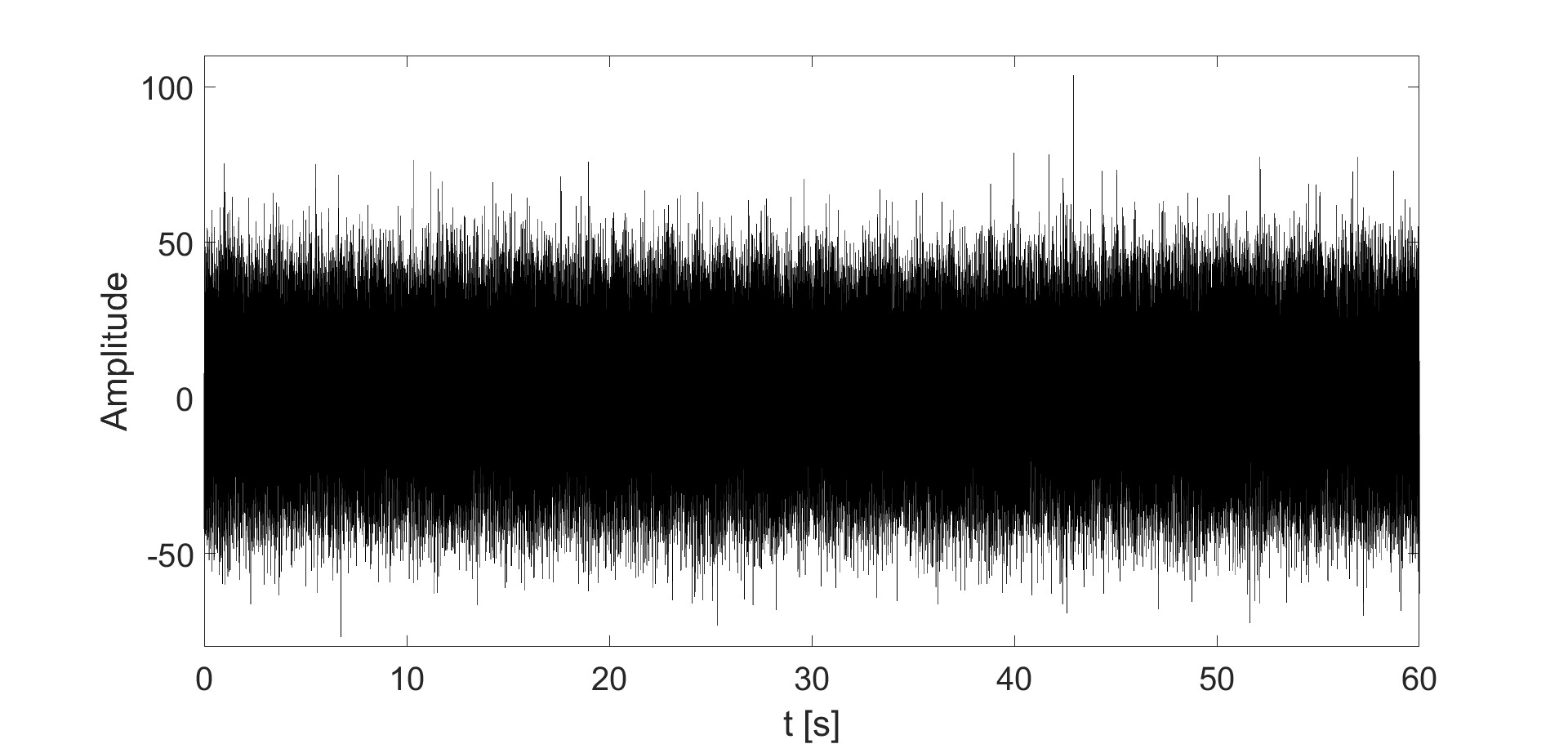}
    \caption{Second example of real vibration signal with minor speed fluctuation (Sig2)}
    \label{Sig2}
\end{figure}

Considered vibrations from industry have 60s duration each. This gives us the possibility to check the results of estimations for larger segment lengths. In the case of the test-rig data, it was 0.5s, 1s, and 2s. For those data, we can additionally consider the 5s and 10s segments. 
Fig. \ref{L62_p1_2012_10_13_line_plot} shows the differences of $f$ for various lengths of segments.

\begin{figure}[H]
\centering
    \includegraphics[width = 16cm]{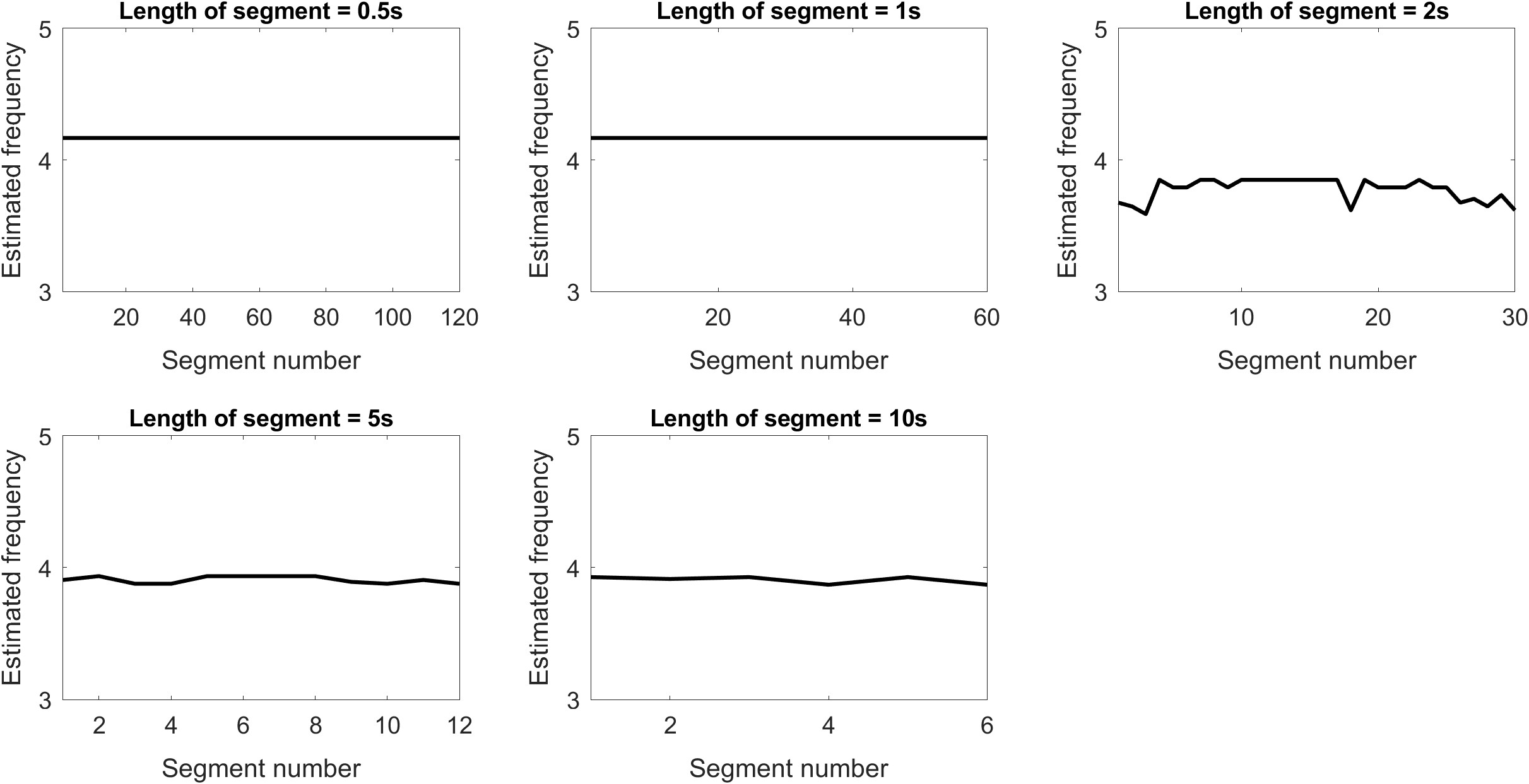}
    \caption{Industrial signal Sig1}  
    \label{L62_p1_2012_10_13_line_plot}
\end{figure}

As we can see, we obtain completely various types of data variability. For small segment lengths such as 0.5s and 1s for a large part of the signal, the estimated frequency was the same. This means that the signal did not change significantly with time. The most interesting is the 2s case. The values change between 3.5 and 4 (for both already mentioned plots, we got values greater than 4). The variance in relation to the mean is relatively high. Most of the estimated $f$ is about 3.8. Only at the beginning, for the 18th segment and at the end of the signal, we have smaller values. For 5s and 10s segments, functions are smoother than for 2s, but oppositely to 1s and 2s they are not constant functions. We have small differences between consecutive values, but the variance is significantly smaller than for 2s segments. Interesting is also the fact that this time again the frequency values are smaller than 4Hz.


The second industrial signal is related to a similar object. Again, the length is similar to the first one, and thus we can also split it into larger segments. The results are shown in Fig. \ref{L215_p3_2016_08_06_line_plot}.

\begin{figure}[H]
\centering
    \includegraphics[width = 16cm]{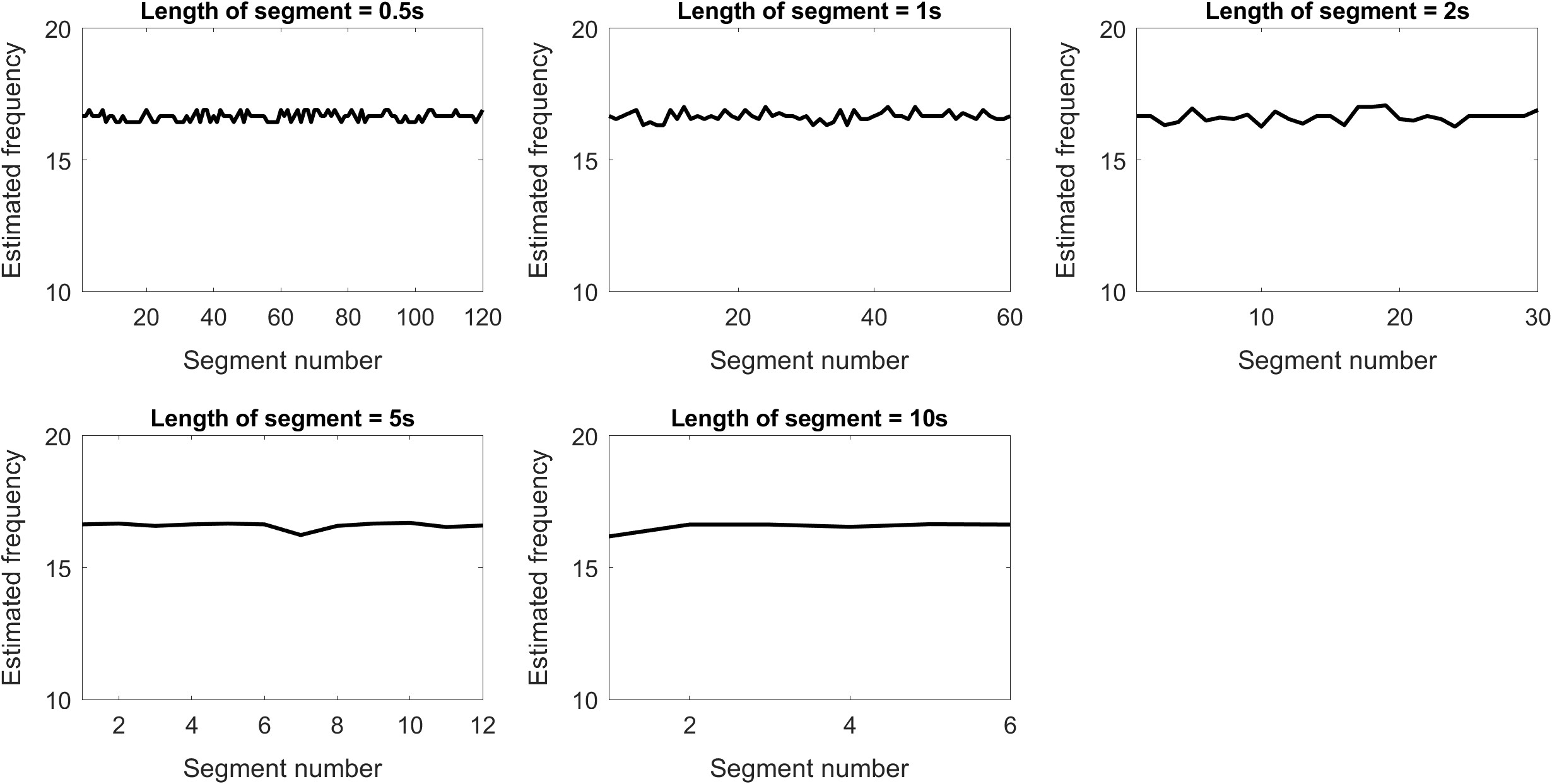}
    \caption{Industrial signal Sig2.}
    \label{L215_p3_2016_08_06_line_plot}
\end{figure}

In Fig. \ref{L215_p3_2016_08_06_line_plot}, we can see that this time for the 0.5s and 1s segments, we do not have constant values. The estimated frequencies oscillate around 17Hz, where in Fig. \ref{L62_p1_2012_10_13_line_plot} it was around 4 Hz (two different frequencies are considered here as the duration of the cycle. They are related to the second and first shafts of a gearbox, respectively). Thus, even if the relative errors could be similar, the absolute errors would be significantly smaller. For both 0.5s and 1s, the estimated values are very similar and do not exceed the range $\left(16,18\right)$. For longer segments, the function becomes smoother. The reason is the lower number of segments.

Based on instantaneous frequency time series, it is hard to 
predict the shape of the distribution of $f$ especially for cases with a smaller number of values. For this purpose, we can look at KDE for 
Sig1 and Sig2, which are shown in Fig. \ref{L62_p1_2012_10_13_denisty_plot} and Fig. \ref{L215_p3_2016_08_06_density_plot}, respectively.

\begin{figure}[H]
\centering
    \includegraphics[width = 16cm]{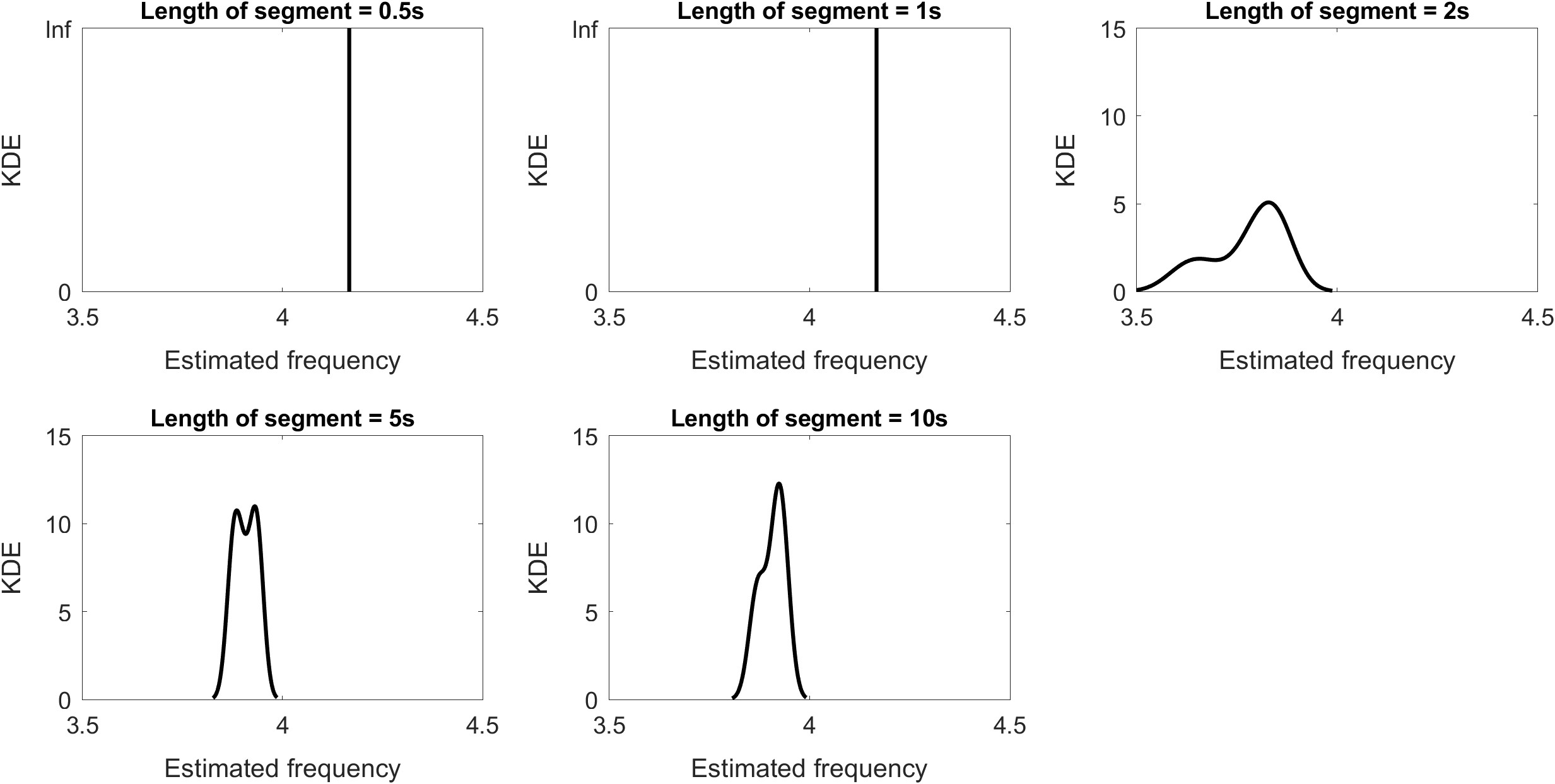}
    \caption{Comparison of KDE for different segment lengths for Sig1}
    \label{L62_p1_2012_10_13_denisty_plot}
\end{figure}

For segments of sizes of 0.5s and 1s (Fig. \ref{L62_p1_2012_10_13_line_plot}) we get the same value for every segment. It is not surprising that KDEs are vertical lines with a maximum value equal to infinity. For the 2s segments, we get the bimodal distribution. The mass is about 3.8. For longer segments (5s, 10s), the mass of the distribution is concentrated in the short frequency range. In case of segment lengths equal to 5s, we also get a bimodal distribution; however, both modes are much closer to each other than for 2s segments. 
The important difference is also that both modes are similar in the context of KDE in opposite to 2s, where the value of mode $\approx3.8$ is clearly larger than the second $\approx3.65$. This means that we may classify the frequency for that case as a uniform distribution. For the plot of the longest segments, shape is similar as for 2s but mass of distribution is concentrated more like for 5s-segments. {As we can see, the relative differences between the estimated values (especially in the context of mass of the distribution) for different segment lengths (e.g. 0.5s and 10s) are very large and can reach 10\%. This means that conclusions based on these results can be very misleading.}
\begin{figure}[H]
\centering
    \includegraphics[width = 16cm]{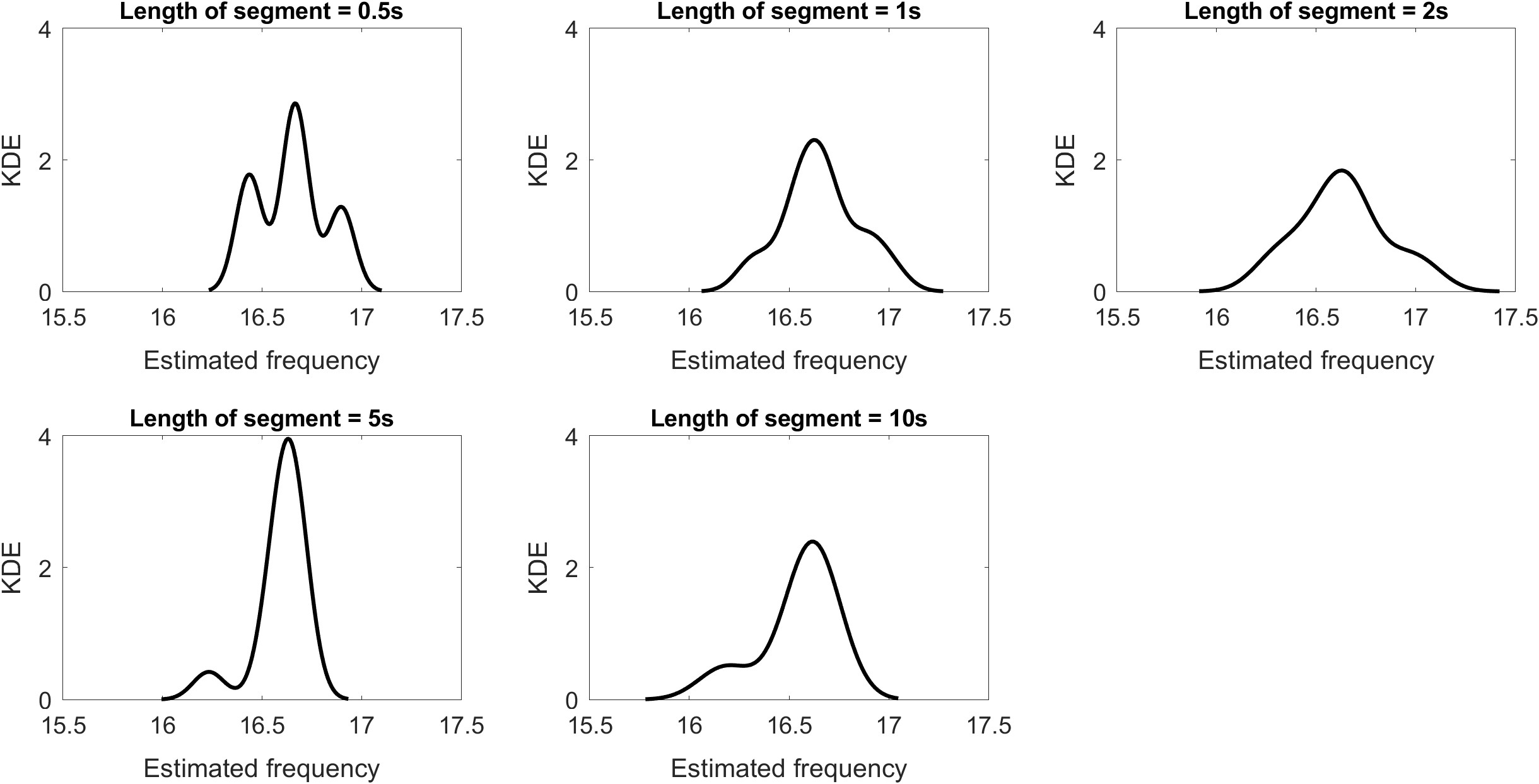}
    \caption{Comparison of KDE for different segment lengths for Sig2}
    \label{L215_p3_2016_08_06_density_plot}
\end{figure}

Finally, we can look at KDE for Sig2. For every plot, the largest values are in a similar place, close to 16.5Hz{, thus relative differences of the estimated values for different segment lengths are smaller than for Sig1}. However, analyzing the signal for different lengths of segments gives us completely different shapes of KDE. For the shortest segments in KDE, we have three modes: 16.45Hz, 16.7Hz, and 16.9Hz. The middle one takes the highest value of KDE. Most of the estimated values are between 16.3 and 17.1 Hz. For the 1s-segments, we get a unimodal distribution. Values vary mainly from 16.1Hz to 17.2Hz. The distribution is almost symmetric. It has a small right skewness. For segments lengths equal to 2s, we have a similar result. It is also a unimodal distribution, and the most frequent value is close to the most frequent value for 1s-segments. However, this time distribution is slightly left-skewed. The fourth and fifth plots are very different from the previous ones. Both are bimodal, but one mode is significantly larger than the second one. Both distributions are left-skewed.

As we could see, we get different shapes of distributions for different considered lengths of segments. For some of them, the range of estimated frequencies was larger, and for others, smaller. It is important to note that one signal can have a relatively large estimated $f$ with a relatively small variance, where the same variance for a significantly smaller estimated $f$ could be large. This means that for the first mentioned case, it is possible to classify frequency as a constant value and, at the same time, for the second case, since frequency is not a constant value. It is hard to decide just by looking at the charts for which one is closer. It can give us general behavior, but not a concrete answer. This is the reason we included a statistical test in our procedure. Interpreting numbers instead of only graphs gives more precise conclusions. In Section \ref{Final decisions} we present the results for real data.

\subsection{Final decisions} \label{Final decisions}

In this section, we present the final identification of frequency variation for test rig Vib2, Sig1, and Sig2. We have two possibilities: we can classify whether the signal has cyclic impulsive signals that appear with constant frequency or not. For the purpose of execute {chi-squared} test, we decided to use the most common significance level: $\alpha=0.05$.

First of all, we analyzed three parts of Vib2. Because each part of the signal is relatively short, only the 0.5s, 1s, and 2s segments were reasonable to consider. For each of these cases, we applied the procedure separately. The results for each important step are shown in Table \ref{table_vib2}.

\begin{table}[H]
    \centering
    \scalebox{0.7}{
    \begin{tabular}{|c|c|c|c|c|}
    \hline
        Part of signal & Segments length & \begin{tabular}{@{}c@{}}Is re-scaled variance \\lower than the threshold?\end{tabular} & {Chi-squared test} & Final classification \\ \hline
        First part & 0.5 s & No & Rejected & Different than a constant value \\ \hline
        First part & 1 s & No & Rejected & Different than a constant value \\ \hline
        First part & 2 s & No & Rejected & Different than a constant value \\ \hline
        Middle part & 0.5 s & Yes & None & Constant value \\ \hline
        Middle part & 1 s & Yes & None & Constant value \\ \hline
        Middle part & 2 s & Yes & None & Constant value \\ \hline
        Third part & 0.5 s & No & Rejected & Different than a constant value \\ \hline
        Third part & 1 s & No & Rejected & Different than a constant value \\ \hline
        Third part & 2 s & No & Rejected & Different than a constant value \\ \hline
    \end{tabular}}
    \caption{\textcolor {black}{A final decision for signal Vib2}}
    \label{table_vib2}
\end{table}

In Table \ref{table_vib2} we can see the results of testing for Vib2. The first column named "Part of signal", as the name suggests, refers to which part of signal is considered in that row. For each of them, we analyzed the length of three segments (second column). In the third column, we checked if re-scaled variance is lower than the threshold. As we can see, for most of the cases we get the answer "No". However, that answer does not mean that the re-scaled variance is significantly greater than the threshold. For this purpose, we have to continue procedure with {chi-squared} test. Only in three rows do we have the answer "Yes". They are for the middle part of Vib2. We could have expected these results based on estimated frequency time series plots and KDE, where for the middle part, for almost every segment, the estimated frequency was the same, and for the first and third parts, different values of estimated $f$ for each of the segments. As already mentioned, if in the third column the answer is "No" then and only then we perform {chi-squared} test. This is the reason why in {"Chi-squared test"} column for the middle part of the signal we have "None". Rest of instances are "Rejected" which means the re-scaled variance is greater than the threshold but not significantly. The last column called "Final classification" shows the classification based on our algorithm. Because we derived thresholds based on simulated signals with constant frequency, thus, if in the fourth column the answer is "Rejected", we classify that the distribution of $f$ is not a constant value. If the answer is "None"  -- distribution of $f$ is classified as a constant value. In Table \ref{table_vib2} only the middle part (when the machine worked stably) has classified the frequency as a constant value, which was intuitive, but not obvious, and worth checking out.

The second signal analyzed is Sig1. As was mentioned, we have a longer signal here; thus we were also able to analyze longer segments: 5s, 10s. Although we have more considered the lengths of the segments, the number of rows in Table \ref{table_L62_p1_2012_10_13} is lower than for Vib2. This is due to the fact that we did not split this signal into parts. This is the reason why the first column of Table \ref{table_vib2} is useless for the final classification of Sig1. We can see this in Table \ref{table_L62_p1_2012_10_13}.

\begin{table}[H]
    \centering
      \scalebox{0.7}{
    \begin{tabular}{|c|c|c|c|}
    \hline
        Segments length & \begin{tabular}{@{}c@{}}Is re-scaled variance \\lower than the threshold?\end{tabular} & {Chi-squared} test & Final classification \\ \hline
        0.5 s & Yes & None & Constant value \\ \hline
        1 s & Yes & None & Constant value \\ \hline
        2 s & No & Rejected & Different than a constant value \\ \hline
        5 s & No & Rejected & Different than a constant value \\ \hline
        10 s & No & Rejected & Different than a constant value \\ \hline
    \end{tabular}}
    \caption{\textcolor{black}{A final decision for Sig1}}
    \label{table_L62_p1_2012_10_13}
\end{table}

As we can see in Table \ref{table_L62_p1_2012_10_13}, for Sig1, similarly to Vib2 more often we get re-scaled variance larger than the threshold, but this time we analyzed the same part of signal. Only the length of the segments was different. That means we can find another signal behavior by considering shorter segments because analyzing segments which are too large can lose some dependencies. On the other hand, it can be caused by a lower resolution for shorter segments, and thus may influence estimation. We have this situation for 0.5s and 1s segments, where re-scaled variance is lower than derived threshold, while for 2s, 5s, and 10s segments it is larger. The same as for Vib2 we perform { chi-squared} test for these longer segments. For all cases where in the second column is "No", we have rejected the hypothesis $H_0$ of the chi-squared} test (equality of variance and the threshold against variance is greater than the threshold). Based on that, we get the final result: for the 0.5s and 1s segments, the frequency is denoted as a constant value and for the 2s, 5s, and 10s segments as different from a constant value.

Finally, we can analyze the last signal -- Sig2. Similarly as for Sig1 shown in Table \ref{table_L62_p1_2012_10_13} we have the same column names -- we did not split the signal into parts and the length of it is also similar to the mentioned signal. This is the reason why we could also analyze longer segments. The results are shown in Table \ref{table_L215_p3_2016_08_06}.

\begin{table}[H]
    \centering
      \scalebox{0.7}{
    \begin{tabular}{|c|c|c|c|}
    \hline
        Segments length & \begin{tabular}{@{}c@{}}Is re-scaled variance \\lower than the threshold?\end{tabular} & Chi-squared test & Final classification \\ \hline
        0.5 s & No & Fail to reject & Constant value \\ \hline
        1 s & No & Rejected & Different than a constant value \\ \hline
        2 s & No & Rejected & Different than a constant value \\ \hline
        5 s & No & Rejected & Different than a constant value \\ \hline
        10 s & No & Rejected & Different than a constant value \\ \hline
    \end{tabular}}
    \caption{\textcolor{black}{A final decision for Sig2}}
    \label{table_L215_p3_2016_08_06}
\end{table}

Totally different results have been obtained for Sig2. As we can see in Table \ref{table_L215_p3_2016_08_06} for every length of segments, the re-scaled variance was larger than the threshold. We also have here a unique decision of the chi-squared test for considered signals: for 0.5s segments, we failed to reject the $H_0$ hypothesis, so even though the variance was greater than the threshold, it was not significantly greater. For all other segments, the $H_0$ hypothesis is rejected. Thus, only for the shortest segment, we classified the frequency as a constant value, but for the rest, the length of the segments as different from a constant value.

\section{Conclusions}
In this paper, the problem of identifying fault frequency variation in the envelope spectrum under speed/load fluctuation is considered. A few real cases have been provided as examples, showing that the problem of a non-stationary condition is not just an academic discussion, but a serious challenge in real applications.

The main factors that influence the estimation of fault frequency have been pointed out (input speed distribution, segment size, SNR). {An influence of jitter in bearings vibration was not considered here, but it will also contribute to frequency variation.} A research task has been defined as automatic, without expert involvement, identification of fault frequency in the envelope of the vibration signal.

A general concept of the transformation of the input speed (with possible variation) into the fault frequency is proposed to illustrate the mechanism of creating uncertainty in the fault frequency estimation.
The deep simulation study confirmed our assumption that constant speed or some minor fluctuation may provide a similar ambiguity that can be hardly acceptable in the case of low-frequency components, complex design, poor SNR, and minor variation of operating conditions.

Thus, the main goal of this paper was to propose a method for identifying the frequency of the fault (that is, detecting machine damage). We proposed to use a statistical perspective, namely, a decision is made based on a set of measurements, as well as statistical testing.
During the simulation study, we have investigated how the above-mentioned factors may influence the final results. 

Finally, we have tested the method on several signals: from a test rig and from industrial machines. Having a tacho signal, we could validate our methods with the variation of the original input speed.
The output of the procedure is a statistical hypothesis test result informing us about presence of speed variation (constant/varying speed). 
If the speed varies, based on the set of signals, we can perform a statistical test to decide if the distribution of $f$ is closer to the uniform or normal distribution. Knowledge about the possible distribution of the fault frequency is important for final, precise fault frequency identification. In case of complex design, low speed machines even minor differences may be important.

\section*{Acknowledgements}
This work is supported by the National Centre for Science under Sheng2 project No. UMO-2021/40/Q/ST8/00024 "NonGauMech - New methods of processing nonstationary signals (identification, segmentation, extraction, modelling) with non-Gaussian characteristics for the purpose of monitoring complex mechanical structures". 

\bibliography{mybibliography}

\begin{thebibliography}{10}
\expandafter\ifx\csname url\endcsname\relax
  \def\url#1{\texttt{#1}}\fi
\expandafter\ifx\csname urlprefix\endcsname\relax\def\urlprefix{URL }\fi
\expandafter\ifx\csname href\endcsname\relax
  \def\href#1#2{#2} \def\path#1{#1}\fi

\bibitem{Zhang2023}
Z.~Zhang, J.~Wang, S.~Li, H.~Bao, B.~Han, Fast nonlinear convolutional sparse filtering: A novel early-stage fault diagnosis method of rolling bearing, Measurement: Journal of the International Measurement Confederation 207 (2023).

\bibitem{capdessus2000cyclostationary}
C.~Capdessus, M.~Sidahmed, J.~Lacoume, Cyclostationary processes: application in gear faults early diagnosis, Mechanical systems and signal processing 14~(3) (2000) 371--385.

\bibitem{combet2009optimal}
F.~Combet, L.~Gelman, Optimal filtering of gear signals for early damage detection based on the spectral kurtosis, Mechanical Systems and Signal Processing 23~(3) (2009) 652--668.

\bibitem{Liu20202323}
D.~Liu, W.~Cheng, W.~Wen, An online bearing fault diagnosis technique via improved demodulation spectrum analysis under variable speed conditions, IEEE Systems Journal 14~(2) (2020) 2323 – 2334, cited by: 13.

\bibitem{Schmidt2021_169}
S.~Schmidt, K.~C. Gryllias, Combining an optimisation-based frequency band identification method with historical data for novelty detection under time-varying operating conditions, Measurement: Journal of the International Measurement Confederation 169~(108517) (2021).

\bibitem{schmidt2020methodology}
S.~Schmidt, A.~Mauricio, P.~S. Heyns, K.~C. Gryllias, A methodology for identifying information rich frequency bands for diagnostics of mechanical components-of-interest under time-varying operating conditions, Mechanical Systems and Signal Processing 142 (2020) 106739.

\bibitem{Sun2021}
R.-B. Sun, F.-P. Du, Z.-B. Yang, X.-F. Chen, K.~Gryllias, Cyclostationary analysis of irregular statistical cyclicity and extraction of rotating speed for bearing diagnostics with speed fluctuations, IEEE Transactions on Instrumentation and Measurement 70~(3514011) (2021) 1--11.

\bibitem{Schmidt2021_158}
S.~Schmidt, P.~S. Heyns, K.~C. Gryllias, An informative frequency band identification framework for gearbox fault diagnosis under time-varying operating conditions, Mechanical Systems and Signal Processing 158 (2021) 107771.

\bibitem{Mauricio2022}
A.~Mauricio, D.~Helm, M.~Timusk, J.~Antoni, K.~Gryllias, Novel cyclo-nonstationary indicators for monitoring of rotating machinery operating under speed and load varying conditions, Journal of Engineering for Gas Turbines and Power 144~(4) (2022).

\bibitem{Mauricio2020_144}
A.~Mauricio, W.~A. Smith, R.~B. Randall, J.~Antoni, K.~Gryllias, Improved envelope spectrum via feature optimisation-gram (iesfogram): A novel tool for rolling element bearing diagnostics under non-stationary operating conditions, Mechanical Systems and Signal Processing 144~(106891) (2020).

\bibitem{Xu2022}
X.~Xu, W.~Li, M.~Zhao, H.~Hu, Mobile device-based bearing diagnostics with varying speeds, Measurement: Journal of the International Measurement Confederation 200 (2022).

\bibitem{borghesani2017cs2}
P.~Borghesani, J.~Antoni, {CS2 analysis in presence of non-Gaussian background noise--Effect on traditional estimators and resilience of log-envelope indicators}, Mechanical Systems and Signal Processing 90 (2017) 378--398.

\bibitem{Wodecki2021}
J.~Wodecki, A.~Michalak, A.~Wyłomańska, R.~Zimroz, {Influence of non-Gaussian noise on the effectiveness of cyclostationary analysis – Simulations and real data analysis}, Measurement: Journal of the International Measurement Confederation 171 (2021) 108814.

\bibitem{cvb}
J.~Hebda-Sobkowicz, R.~Zimroz, M.~Pitera, A.~Wy{\l}oma{\'n}ska, {Informative frequency band selection in the presence of non-Gaussian noise--a novel approach based on the conditional variance statistic with application to bearing fault diagnosis}, Mechanical Systems and Signal Processing 145 (2020) 106971.

\bibitem{hebda2020selection}
J.~Hebda-Sobkowicz, R.~Zimroz, A.~Wy{\l}oma{\'n}ska, {Selection of the Informative Frequency Band in a Bearing Fault Diagnosis in the Presence of Non-Gaussian Noise —- Comparison of Recently Developed Methods}, Applied Sciences 10~(8) (2020) 2657.

\bibitem{randall2011rolling}
R.~B. Randall, J.~Antoni, Rolling element bearing diagnostics—a tutorial, Mechanical Systems and Signal Processing 25~(2) (2011) 485--520.

\bibitem{Rai2016289}
A.~Rai, S.~Upadhyay, A review on signal processing techniques utilized in the fault diagnosis of rolling element bearings, Tribology International 96 (2016) 289 – 306.

\bibitem{Zhao2022}
Q.~Zhao, J.~Wang, J.~Yin, P.~Zhang, Z.~Xie, Peak envelope spectrum fourier decomposition method and its application in fault diagnosis of rolling bearings, Measurement: Journal of the International Measurement Confederation 198 (2022).

\bibitem{Urbanek_Jacek_Measurement_2011}
J.~Urbanek, T.~Barszcz, R.~Zimroz, W.~Bartelmus, F.~Millioz, N.~Martin, Measurement of instantaneous shaft speed by advanced vibration signal processing - application to wind turbine gearbox, Metrology and Measurement Systems~(No 4) (2011) 701--712.

\bibitem{Hou2021}
F.~Hou, I.~Selesnick, J.~Chen, G.~Dong, Fault diagnosis for rolling bearings under unknown time-varying speed conditions with sparse representation, Journal of Sound and Vibration 494 (2021).

\bibitem{Wang2019391}
T.~Wang, F.~Chu, Bearing fault diagnosis under time-varying rotational speed via the fault characteristic order (fco) index based demodulation and the stepwise resampling in the fault phase angle (fpa) domain, ISA Transactions 94 (2019) 391 – 400.

\bibitem{Wang2019}
L.~Wang, J.~Xiang, Y.~Liu, A time-frequency-based maximum correlated kurtosis deconvolution approach for detecting bearing faults under variable speed conditions, Measurement Science and Technology 30~(12) (2019).

\bibitem{Ming2016367}
A.~Ming, W.~Zhang, Z.~Qin, F.~Chu, Fault feature extraction and enhancement of rolling element bearing in varying speed condition, Mechanical Systems and Signal Processing 76-77 (2016) 367 – 379.

\bibitem{Borghesani201323}
P.~Borghesani, R.~Ricci, S.~Chatterton, P.~Pennacchi, A new procedure for using envelope analysis for rolling element bearing diagnostics in variable operating conditions, Mechanical Systems and Signal Processing 38~(1) (2013) 23 – 35.

\bibitem{Zhao2016109}
D.~Zhao, J.~Li, W.~Cheng, W.~Wen, Compound faults detection of rolling element bearing based on the generalized demodulation algorithm under time-varying rotational speed, Journal of Sound and Vibration 378 (2016) 109 – 123.

\bibitem{Zhao2015}
D.~Zhao, J.~Li, W.~Cheng, Feature extraction of faulty rolling element bearing under variable rotational speed and gear interferences conditions, Shock and Vibration 2015 (2015).

\bibitem{Chaari_2012_SV}
F.~Chaari, W.~Bartelmus, R.~Zimroz, T.~Fakhfakh, M.~Haddar, Gearbox vibration signal amplitude and frequency modulation, Shock and Vibration~(No 4) (2012) 635--652.

\bibitem{Kruczek_poisson}
P.~Kruczek, M.~Polak, A.~Wyłomańska, W.~Kawalec, R.~Zimroz, Application of compound poisson process for modelling of ore flow in a belt conveyor system with cyclic loading, International Journal of Mining, Reclamation and Environment 32~(6) (2018) 376--391.

\bibitem{COMBET20091382}
F.~Combet, R.~Zimroz, A new method for the estimation of the instantaneous speed relative fluctuation in a vibration signal based on the short time scale transform, Mechanical Systems and Signal Processing 23~(4) (2009) 1382--1397.

\bibitem{COCCONCELLI2012667}
M.~Cocconcelli, L.~Bassi, C.~Secchi, C.~Fantuzzi, R.~Riccardo, An algorithm to diagnose ball bearing faults in servomotors running arbitrary motion profiles, Mechanical Systems and Signal Processing 27 (2012) 667--682.

\bibitem{ref1}
C.~Sammut, G.~I. Webb (Eds.), Mean Squared Error, Springer US, Boston, MA, 2010, pp. 653--653.

\bibitem{kde}
Y.-C. Chen, A tutorial on kernel density estimation and recent advances, Biostatistics \& Epidemiology 1 (04 2017).

\bibitem{stat2}
G.~Snedecor, W.~Cochran, Statistical Methods. 6th Edition, Iowa State University Press, 1983.

\bibitem{forbes2010statistical}
C.~Forbes, M.~Evans, N.~Hastings, B.~Peacock, Statistical Distributions, Wiley Series in Probability and Statistics - Applied Probability and Statistics Section Series, Wiley, 2010.

\bibitem{erf}
M.~Abramowitz, I.~Stegun, Handbook of Mathematical Functions, Dover, New York, 1965.

\bibitem{SIJBERS19961157}
J.~Sijbers, P.~Scheunders, N.~Bonnet, D.~{Van Dyck}, E.~Raman, Quantification and improvement of the signal-to-noise ratio in a magnetic resonance image acquisition procedure, Magnetic Resonance Imaging 14~(10) (1996) 1157--1163.

\end{thebibliography}


\end{document}